\documentclass{aa}  
\usepackage{graphicx}
\usepackage{multirow}
\usepackage{amssymb}
\usepackage{xcolor}
\usepackage{txfonts}
\usepackage{hyperref}
\usepackage{natbib}
\bibpunct{(}{)}{;}{a}{}{,}
\usepackage{lipsum, babel}

\begin{document} 

   \title{The dense and non-homogeneous circumstellar medium revealed in radio wavelengths around the Type Ib SN\,2019oys}
   
   \subtitle{}

   \author{I. Sfaradi
          \inst{1}\fnmsep\thanks{\email{ itai.sfaradi@mail.huji.ac.il}}
          \and
          A. Horesh\inst{1}
          \and
          J. Sollerman\inst{2}
          \and 
          R. Fender\inst{3}
          \and
          L. Rhodes\inst{3}
          \and
          D. R. A. Williams\inst{4}
          \and
          J. Bright\inst{3}
          \and
          D. A. Green\inst{5}
          \and
          S. Schulze\inst{6,7}
          \and
          A. Gal-Yam\inst{8}
          }

   \institute{Racah Institute of Physics. The Hebrew University of Jerusalem. Jerusalem 91904, Israel
   \and
   Department of Astronomy, The Oskar Klein Center, Stockholm University, AlbaNova, SE-10691 Stockholm, Sweden
   \and
   Astrophysics, Department of Physics, University of Oxford, Keble Road, Oxford OX1 3RH, UK
   \and
   Jodrell Bank Centre for Astrophysics, School of Physics and Astronomy, The University of Manchester, Manchester, M13 9PL, UK
   \and
   Astrophysics Group, Cavendish Laboratory, 19 J. J. Thomson Ave., Cambridge CB3 0HE, UK
   \and
   Center for Interdisciplinary Exploration and Research in Astrophysics (CIERA), Northwestern University, 1800 Sherman Ave., Evanston, IL 60201, USA
   \and
   The Oskar Klein Centre, Department of Physics, Stockholm University, Albanova University Center, 106 91 Stockholm, Sweden
   \and
   Department of Particle Physics and Astrophysics, Weizmann Institute of Science, 234 Herzl St, 7610001 Rehovot, Israel
   }

   \date{Submitted \today}

  \abstract
   {Mass loss from massive stars, especially towards the end of their lives, plays a key role in their evolution. Radio emission from core-collapse supernovae (SNe) serves as a probe of the interaction of the SN ejecta with the circumstellar medium (CSM) and can reveal the mass-loss history of the progenitor.}
   {We present here broadband radio observations of the CSM interacting SN\,2019oys. SN\,2019oys was first detected in the optical and was classified as a Type Ib SN. Then, about $\sim 100$ days after discovery, it showed an optical rebrightening and a spectral transition to a spectrum dominated by strong narrow emission lines, which suggests strong interaction with a distant, dense, CSM shell.}
   {We modeled the broadband, multi-epoch, radio spectra, covering $2.2$ to $36$ GHz and spanning from $22$ to $1425$ days after optical discovery, as a synchrotron emitting source. Using this modeling we characterized the shockwave and the mass-loss rate of the progenitor.}
   {Our broadband radio observations show strong synchrotron emission. This emission, as observed $201$ and $221$ days after optical discovery, exhibits signs of free--free absorption from the material in front of the shock traveling in the CSM. In addition, the steep power law of the optically thin regime points towards synchrotron cooling of the radiating electrons. Analyzing these spectra in the context of the SN-CSM interaction model gives a shock velocity of $14,000 \, \rm km\,s^{-1}$, and an electron number density of $2.6 \times 10^{5} \, \rm cm^{-3}$ at a distance of $2.6 \times 10^{16} \, \rm cm$. This translates to a high mass-loss rate from the progenitor massive star of $6.7 \times 10^{-4} \, \rm M_{\odot}\,yr^{-1}$ for an assumed wind of $100 \, \rm km\,s^{-1}$ (assuming constant mass-loss rate in steady winds). The late-time radio spectra, $392$ and $557$ days after optical discovery, are showing broad spectral peaks. We show that this can be explained by introducing a non-homogeneous CSM structure.}
   {}


    \authorrunning{Sfaradi et al.}
    \titlerunning{A non-homogenous CSM shell around SN\,2019oys}

   \maketitle

\section{Introduction} 
\label{sec:intro}

Mass-loss from massive stars, stars with $M_{ZAMS} \geq 8 \: \rm{M_{\odot}}$ ($M_{ZAMS}$ - Zero-Age Main Sequence mass of the star), has a great impact on their luminosity, lifetime, composition and mass (e.g., \citealt{Langer_2012}). The mass lost from the massive star, either via winds or other mass-loss mechanisms (e.g., violent mass eruptions, binary interaction), forms a circumstellar material (CSM) surrounding the massive star. Towards the end of its life, before the massive star explodes as a core-collapse supernova (CCSN), mass-loss can heavily influence the fate of the star and the resulting supernova (SN; e.g., \citealt{smith_2014}). Despite its importance, mass-loss at this last stage of stellar evolution is still highly unconstrained empirically.

Observations of CCSNe across the electromagnetic spectrum are showing a wide variety of properties. In the optical, distinct features in the SN spectral lines are used to classify CCSNe to sub-types \citep{Filippenko_1997}. Hydrogen-rich SNe, such as Type II, are considered to be a result of the explosion of a massive star that retained its hydrogen envelope in its final stages. Stripped envelope SNe (Types Ib and Ic) on the other hand, lack hydrogen spectral features, and are the result of massive stars that lost their hydrogen envelopes (and in some cases even their helium envelope) in some sort of enhanced mass loss at the end of their lifes \citep{SN_handbook}.

Radio observations of CCSNe probe the interaction between the SN ejecta and the CSM around the SN progenitor star \citep{chevalier_radio_1982}, and thus play a key role in revealing the mass-loss history of massive stars. While it is well established that the emission in radio from CCSNe is synchrotron emission that originates at the interaction region \citep{chevalier_1981, chevalier_fransson_2006, weiler_2002}, there are variations between different CCSNe in the observed radio emission. For example, most SNe exhibit clear self-absorbed synchrotron radio spectra \citep{chevalier_1998}, however, some SNe (e.g. SN\,1979C; \citealt{weiler_1991}, SN\,1993J; \citealt{weiler_2007}) show steep power-laws in the optically thick regime which point towards a spectrum associated with free--free absorption (FFA). This implies a more dense CSM around the massive progenitor star. Furthermore, steep power laws in the optically thin regime can arise from electron cooling either by synchrotron and/or inverse Compton cooling (e.g. SN\,2012aw; \citealt{yadav_2014}, SN\,2013df; \citealt{kamble_2016}, SN\,2020oi; \citealt{horesh_2020}, SN\,2016X; \citealt{SN2016x_2022}).

While it is common to assume a spherical, homogeneous, CSM structure, deviations from this simple scenario have been observed in several CCSNe. Some SNe (e.g. SN\,2014C; \citealt{anderson_2017}) exhibit a double peak in their radio light curve suggesting that the SN ejecta interact with two separate CSM shells deposited in two separate mass-loss stages. Equipartition analysis of the radio spectral peaks of SN\,2004C \citep{demarchi_2022} suggested a shock wave that initially interacts with an inner, relatively flat, density profile, and later, with an outer $\sim r^{-2}$ profile. Panchromatic, optical and radio, observations of the Type Ib SN\,2019tdf \citep{SN2019tsf} point towards a scenario of a SN ejecta interacting with spherically-asymmetric CSM, potentially disk-like. Some SNe exhibit broader radio spectral peaks. These broad peaks are associated with CSM inhomogeneities caused by variations in the distribution of relativistic electrons and/or the magnetic field strength within the synchrotron source. Radio observations of such broad peaks (e.g. SN\,2003L; \citealt{soderberg_2005}, PTF11qcj; \citealt{Bjornsson_2017}, Master OT J120451.50+265946.64; \citealt{Poonam_2019}) shed light on variable mass-loss processes in the massive star finale stages. Such variations from the typical synchrotron spectra and light curve can reveal deviation from the simple spherical CSM structure around the progenitor (which is the standard assumption) and can be caused, for example, by variable mass loss from the progenitor during its evolution.

Here, we present broadband radio observations of SN\,2019oys. SN\,2019oys (a.k.a. ZTF19abucwzt) was first detected on August 28, 2019 (JD = 2458723.98), with the Palomar Schmidt 48-inch (P48) Samuel Oschin telescope as part of the Zwicky Transient Facility (ZTF) survey (\citealt{Bellm_2019}; \citealt{Graham_2019}; \citealt{sollerman_2020}). The SN was reported to the Transient Name Server (TNS\footnote{https://www.wis-tns.org}) on August 29, 2019, with the first detection of a host-subtracted magnitude of $19.14 \pm 0.12$ mag (in g-band). It is positioned in the spiral galaxy CGCG 146-027 NED01 and the J2000.0 coordinates of SN\,2019oys are $\alpha = 07^{\rm h} 07^{\rm m} 59.26^{\rm s}$, $\delta = +31^{\rm \circ}$39'55.3". The distance to the source (based on its redshift, $z=0.0165$; \citealt{sollerman_2020}) is $73.2 \pm 3.7$ Mpc (where we adopted $5\%$ uncertainty on this distance based on the distances published in the NED\footnote{https://ned.ipac.caltech.edu} catalog). SN\,2019oys was discovered during the decline of the optical light curve and was first classified as Type Ib SN. Then, about $100$ days after optical discovery, it showed a re-brightening in the $g$ and $r$ bands. Deep Keck spectra \citep{sollerman_2020} obtained at the re-brightening phase revealed a plethora of narrow high-ionization lines, including coronal lines. The evolution of the optical light curve and the makeover of the spectrum suggests an onset of strong CSM interaction at late times ($\sim 100$ days after optical discovery).

The paper is structured as follows: in \S\ref{sec: radio_obs} we present our comprehensive radio observations of SN\,2019oys conducted with the Karl G. Jansky Very Large Array (VLA) and the Arcminute Microkelvin Imager -- Large Array (AMI-LA). In \S\ref{sec: modeling_and_analysis} we model the first two radio spectra as arising from a blastwave interaction with a spherical and homogeneous CSM structure. We also consider electron cooling effects and analyze the temporal evolution of the detailed $15.5$ GHz light curve. In \S\ref{sec: inhomogen} we discuss the indications of a non-homogeneous CSM structure seen in the late-time radio spectra. Our conclusions are discussed in \S\ref{sec: conclusions}. 

\section{Radio observations}
\label{sec: radio_obs}

As part of our regular monitoring campaign for CCSNe, we first observed the field of SN\,2019oys in radio wavelengths with the AMI-LA \citep{zwart_2008,hickish_2018} on September 19, 2019, about 22 days after optical discovery. We detected a point source at the position of SN\,2019oys with a flux level of $0.35$\,mJy at a central frequency of $15.5$\,GHz. Follow-up observations in the $15.5$\,GHz band were conducted using AMI-LA, and multi-wavelength observations were performed using the Karl G. Jansky Very Large Array (VLA).

\subsection{The Arcminute Microkelvin Imager - Large Array}
\label{subsec: AMI-LA}

AMI-LA is a radio  interferometer comprised of eight 12.8 m diameter antennas producing 28 baselines that extend from 18 m up to 110 m in length and operate with a 5 GHz bandwidth, divided into eight channels, around a central frequency of 15.5 GHz. This results in a synthesized beam of $\sim 30$\,arcsec.

Observations of SN\,2019oys were initially reduced, flagged and calibrated using $\tt{reduce \_ dc}$, a customized AMI-LA data reduction software package (\citealt{perrott_2013}). Phase calibration was conducted using short interleaved observations of J0823+2223, while daily observations of 3C286 were used for absolute flux calibration. Additional flagging was performed using the Common Astronomy Software Applications (CASA; \citealt{2007ASPC..376..127M}). Images of the field of SN\,2019oys were produced using CASA task CLEAN in an interactive mode. We fitted the source in the phase center of the images with the CASA task IMFIT, and calculated the image rms with the CASA task IMSTAT. We estimate the error of the peak flux density to be a quadratic sum of the error produced by CASA task IMFIT and $10$\,\% calibration error. The flux density at each time is reported in Table~\ref{tab:Radio_Observations} in Appendix \ref{sec: tables}.

\subsection{The Karl G. Jansky Very Large Array}
\label{subsec: VLA}

We observed the field of SN\,2019oys with the VLA on several epochs starting January 10, 2020. The observations (under our DDT programs VLA/20A-421 and VLA/21A-394; PI Horesh) were performed in the S- ($3$\,GHz), C- ($5$\,GHz), X- ($10$\,GHz), Ku- ($15$\,GHz), K- ($22$\,GHz), and Ka- ($33$\,GHz) bands. The VLA was in C configuration during the first and second observations on March 16 and April 5, 2020, in the more extended B configuration in the third observation on September 23, 2020, and in the hybrid $\rm A\rightarrow D$ configuration on March 7, 2021. 

We calibrated the data using the automated VLA calibration pipeline available in the CASA package. Additional flagging was conducted manually when needed. Our primary flux density calibrator was 3C286, while J1215+1654 was used as a phase calibrator. Images of the SN\,2019oys field were produced using the CASA task CLEAN in an interactive mode. Each image was produced using data from within the VLA bands, resulting in different spectral resolutions for the different bands. We also produced images of the full band data for each epoch. 

Our observations showed a source at the phase center (besides the S-band observations of the first two epochs), which we fitted with the CASA task IMFIT. The image rms was calculated using the CASA task IMSTAT. A summary of the flux density at different observing times and frequencies, for the full band images, are reported in Table~\ref{tab:Radio_Observations} in Appendix \ref{sec: tables}. We estimate the error of the peak flux density to be a quadratic sum of the error produced by CASA task IMFIT, and $10$\,\% calibration error.

\subsection{The Very Large Array Sky Survey}
\label{subsec: VLASS}
In addition to the observations conducted by us as reported above, the field of view around the SN position was covered, with a central frequency of $3$ GHz, by the VLA under the initiative of the Very Large Array Sky Survey (VLASS; \citealt{vlass}) on 2019, April 22 (epoch 1.2), and 2021, December 3 (epoch 2.2). Examining the cutout images using the Canadian Initiative for Radio Astronomy Data Analysis (CIRADA\footnote{http://cutouts.cirada.ca}) reveals no point source emission at the position of the SN on epoch 1.2, about four months prior to the optical discovery with $3\sigma$ rms of $0.39$ mJy, and thus set a limit on the possible contamination of the SN emission from a pre-existing radio source. On the other hand, the image obtained on epoch 2.2, $823$ days after optical discovery, shows a point source with a flux level of $3.4 \pm 0.4$ mJy (assuming $10 \%$ calibration error).

\section{Modeling of the radio data - a spherical and homogeneous CSM structure}
\label{sec: modeling_and_analysis}

In the simple (spherical and homogeneous CSM structure) SN-CSM interaction model (\citealt{chevalier_1981}; \citealt{chevalier_1998}), the radio emission observed from a SN is attributed to the interaction between the SN ejecta and the CSM. This plowing of the SN ejecta into the CSM generates a shock wave. In the shock front, electrons are accelerated to relativistic velocities, with a power-law energy distribution ($ N = N_0 E^{-p}$; where $N$ is the density of relativistic electrons per unit energy, $N_0$ is a constant, and $p$ is the power law index), and magnetic fields are enhanced. A fraction of the shockwave energy is deposited in relativistic electrons ($\epsilon_{\rm e}$), and a fraction of the shockwave energy is converted to magnetic fields ($\epsilon_{\rm B}$). The relativistic electrons are gyrating in the presence of the magnetic fields, and radiate the synchrotron emission observed in radio wavelengths. Part of the emission can be absorbed by either synchrotron self-absorption (SSA; \citealt{chevalier_1998}) and/or FFA \citep{weiler_2002}. The synchrotron self-absorbed spectrum exhibits an optically thick regime of $F_{\rm \nu} \propto \nu^{5/2}$, a radio spectral peak, and an optically thin regime of $F_{\rm \nu} \propto \nu^{-(p-1)/2}$ (assuming no electron cooling). For electrons that are gyrating in a thin shell with volume filling factor $f$ at a radius $R$, in the presence of a constant magnetic field, $B$, the full spectral shape is described in \cite{chevalier_1998}.

A typical assumption is that the magnetic energy density, $B^2/8 \pi$, is a fraction, $\epsilon_{\rm B}$, of the post-shock energy, $\sim \rho v_{\rm sh}^2$. For a CSM that was deposited by constant mass-loss from the progenitor via a steady wind with velocity $v_{\rm w}$, the density structure is $\rho \sim \dot{M}/ \left( v_{\rm w} r^2 \right)$, where $\dot{M}$ is the mass-loss rate. Assuming a constant shock velocity, $v_{\rm sh} = R/\Delta t$, gives
\begin{align}
    \label{eq: mass-loss}
    \nonumber \dot{M} = \, & 5.2 \times 10^{-7} \, \left( \frac{\epsilon_{\rm B}}{0.1} \right)^{-1} \left( \frac{B}{1 \, \rm{G}} \right)^{2} \left( \frac{\Delta t}{10 \, \rm{Day}} \right)^{2} \\
    & \left( \frac{v_{\rm w}}{100 \, \rm{km\,s^{-1}}} \right) \, \rm M_{\odot}\,yr^{-1}.
\end{align}

External FFA can take place as discussed in many cases (e.g., \citealt{weiler_2002,chevalier_fransson_2006,horesh_2012,nayana_2018}) and was observed for several SNe (e.g., SN\,1979C; \citealt{weiler_1991}, SN\,1993J; \citealt{weiler_2007}, SN\,2013df; \citealt{kamble_2016}). This attenuates the optically thick regime of the synchrotron self-absorbed spectrum by a factor of $e^{-\tau_{\rm ff}}$ where the optical depth for FFA (see e.g., \citealt{horesh_2012}) is
\begin{align}
    \label{eq: ff_od1}
    \tau_{\rm ff} = 1.7 \times 10^{-10} \, \left( \frac{T_{\rm e}}{10^5 \, \rm{K}} \right)^{-1.35} \left( \frac{\nu}{5 \, \rm{GHz}} \right)^{-2.1} \rm{EM},
\end{align}
where $\rm T_{\rm e}$ is the temperature of the electrons in the CSM, and EM is the integral of the electron density along the line of sight in units of $\rm cm^{-6}pc$. Assuming a spherical, $r^{-2}$, density profile gives
\begin{align}
    \label{eq: EM}
    {\rm EM} = \int_{r_*} ^{\infty} n^2 \left(r\right) \, {\rm d}r = \frac{1}{3} n_* ^2 r_*,
\end{align}
where $n_*$ is the density of the electrons at $r_*$. Substituting into Eq.\ref{eq: ff_od1}, and assuming constant shock velocity gives\footnote{We note that the assumption of a spherical, $r^{-2}$, density profile arises from the assumption of a constant mass-loss rate from steady winds which is not always true. In addition, the EM assumes a wide shell such that the integral of the electron density along the line of sight (Eq. \ref{eq: EM}) does not depend on its width. If the interacting CSM shell is relatively thin and/or varies from the simple spherical density profile, as discussed later in \S\ref{sec: inhomogen}, FFA optical depth may vary.}
\begin{align}
    \label{eq: ff_od}
    \nonumber \tau_{\rm ff} = & \, 0.76 \, \left( \frac{\dot{M} \left[ 10^{-6} \, \rm{M_{\odot}\,yr^{-1}} \right]}{v_{\rm w} \left[ 10 \, \rm{km\,s^{-1}} \right]} \right)^{2} \left(\frac{T_{\rm e}}{10^5 \, \rm{K}} \right)^{-1.35} \\ & \left(\frac{v_{\rm sh}}{10^4 \, \rm{km\,s^{-1}}} \right)^{-3} \left(\frac{\nu}{5 \, \rm{GHz}} \right)^{-2.1} \left(\frac{\Delta t}{10 \, \rm{Day}} \right)^{-3}.
\end{align}

\subsection{Our radio dataset}
\label{subsec: radio_dataset}

\begin{figure*}
\includegraphics[width=1\linewidth]{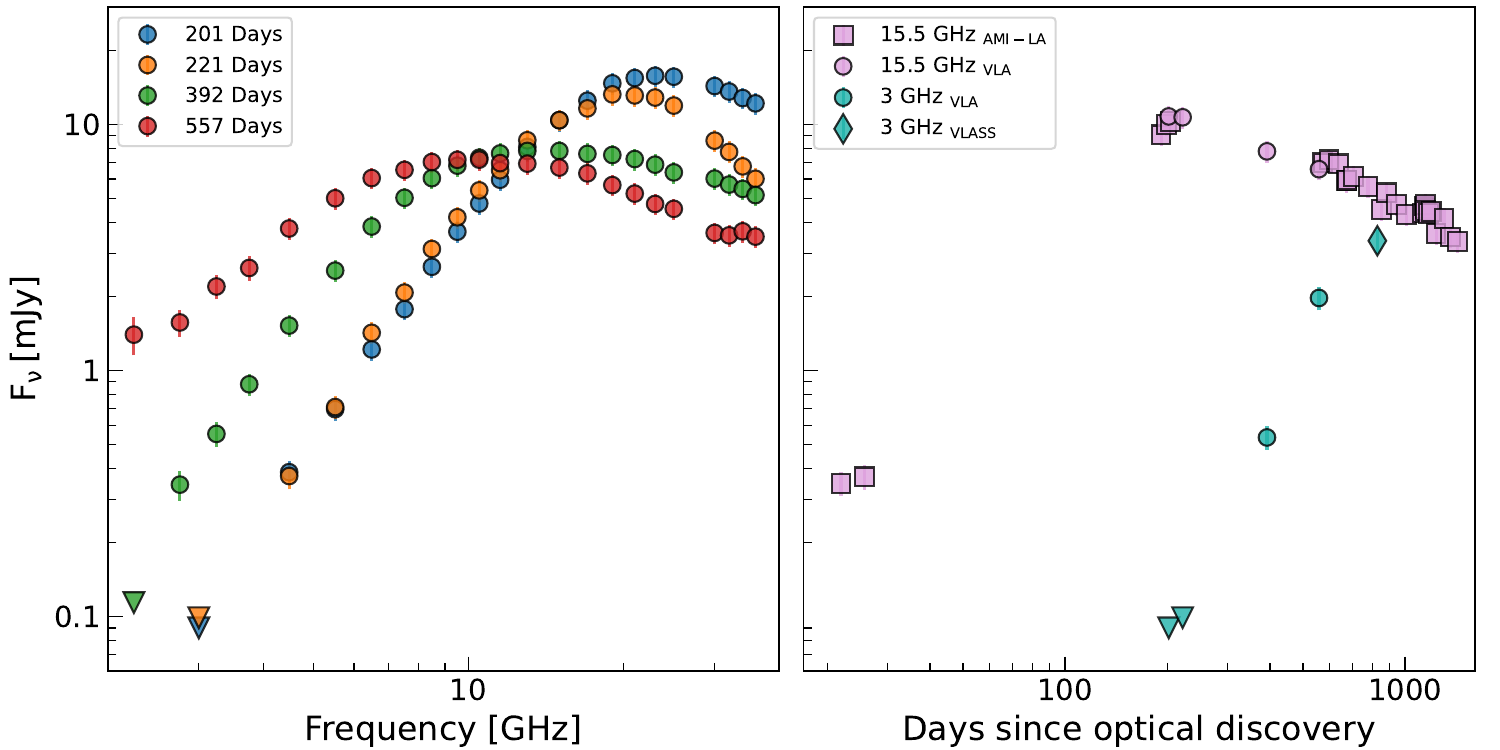}
\caption{\footnotesize{Radio observations of SN\,2019oys with the VLA and AMI-LA (see \S\ref{sec: radio_obs}). The left panel is showing the broadband spectra obtained with the VLA in this work on four different epochs. The right panel is showing the $15.5$ GHz light curve obtained with AMI-LA (square markers), and the VLA (circle markers), and a $3$ GHz light curve obtained by our VLA (circle markers) and VLASS (diamond marker) observations. The $3\sigma$ rms upper limits obtained by the VLA are marked with triangles. \label{fig: VLA_AMI}}}
\end{figure*}

Early radio observations of SN\,2019oys, taken with AMI-LA on $\Delta t = 22$ and $26$ days (where $\Delta t$ is the time in days after the optical discovery), show a low emission level of $\sim 0.3-0.4$ mJy at $15.5$ GHz (see right panel of Fig. \ref{fig: VLA_AMI}). After a gap of about six months in observations, we triggered AMI-LA following the late-time CSM interaction observed in the optical. The flux measured from the SN at the same frequency had increased significantly to a level of $\sim 10$ mJy. The spectrum obtained with the VLA on $\Delta t = 201$ days revealed a broadband optically thick emission up to the spectral peak at $\sim 20$ GHz. However, due to the lack of spectral coverage at high frequencies ($\geq 36$ GHz), we do not detect the optically thin emission to its full extent at that epoch. A full optically thick to thin spectrum is observed on $\Delta t = 221$ days. Two additional broadband VLA spectra obtained on $\Delta t = 392$ and $557$ days are exhibiting extremely broad spectral peaks (see left panel of Fig. \ref{fig: VLA_AMI}). Such broad peaks cannot be explained by a shockwave interaction with a spherical and homogenous CSM structure.

Thus, in \S\ref{subsec: simple_sync} we analyze the early spectra (on $\Delta t = 201$ and $221$ days) by assuming they arise from an SN ejecta interacting with a simple homogeneous CSM shell, as described above in the SN-CSM model. We then discuss the possibility of electron cooling (namely synchrotron cooling) and its effect on the physical parameters (i.e., $R$, $B$, and as a result, $v_{\rm sh}$ and $\dot{M}$) in \S\ref{subsec: cooling}. We examine the atypical, slow, temporal evolution of SN\,2019oys through its detailed $15.5$ GHz observations with AMI-LA (see \S\ref{subsec: temporal}). We discuss the analysis of the late-time spectra (on $\Delta t = 392$ and $557$ days) showing broad spectral peaks in \S\ref{sec: inhomogen}.

\subsection{Single epoch modeling of the synchrotron spectra}
\label{subsec: simple_sync}

The functional form of a synchrotron self-absorbed spectrum can be described by its observed peak flux density, $F_{\nu,a}$, the synchrotron self absorption frequency (at which the spectrum peaks), $\nu_a$, and the power-law index of the optically thin regime, $\beta$, \citep{chevalier_1998,weiler_2002}:
\begin{align}
    F_{\nu} = 1.582 \times F_{\rm \nu,a} \left( \frac{\nu}{\nu_{\rm a}} \right)^{\frac{5}{2}} \left( 1 - \exp \left[-\left( \frac{\nu}{\nu_{\rm a}} \right)^{-\left( \frac{5}{2} + \beta \right)} \right]\right).
    \label{eq: parameterized}
\end{align}
This two-power-law spectrum will be attenuated, in the case of external FFA, by a factor of
\begin{align}
    \label{eq: ff_od_param}
    \exp \left[ -\left(\frac{\nu}{\nu_{\rm ff}}\right)^{-2.1} \right],
\end{align}
where $\nu_{\rm ff}$ is the FFA frequency which can be defined by Eq. \ref{eq: ff_od}.

\begin{figure*}
\includegraphics[width=0.96\linewidth]{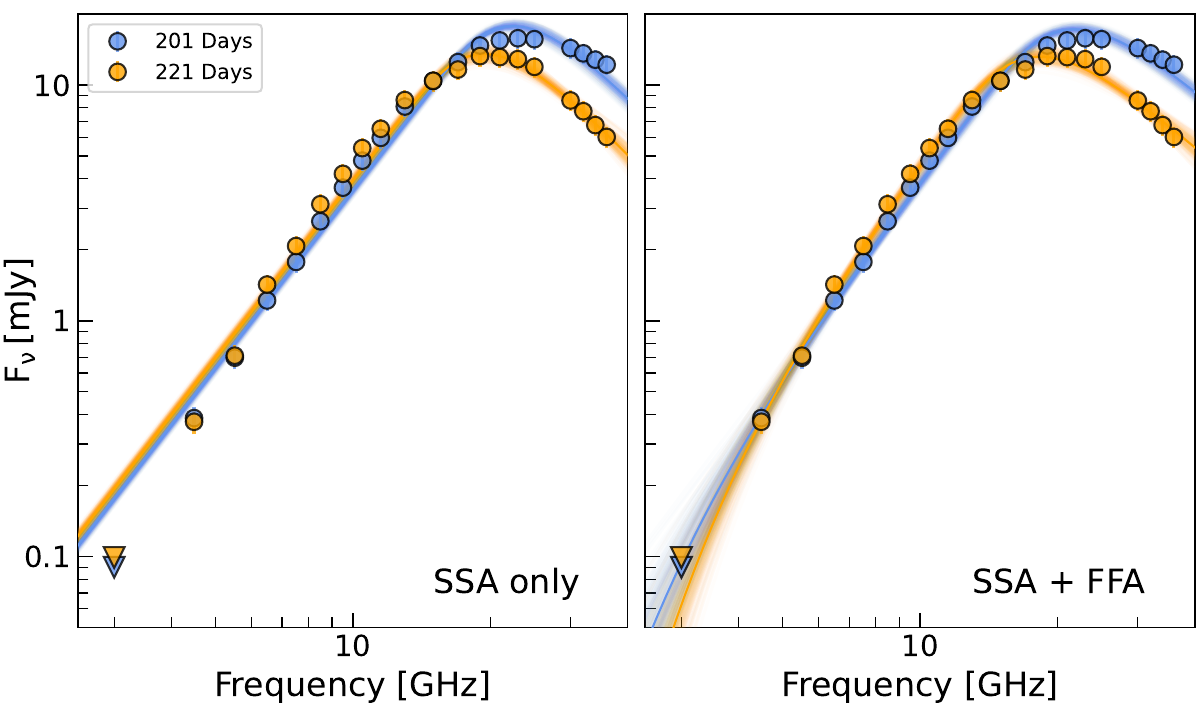}
\caption{\footnotesize{Modeling of the radio spectra obtained by the VLA on $\Delta t = 201$ and $221$ days after optical discovery. The left panel is showing the fitting of Eq. \ref{eq: parameterized} to the spectra, assuming SSA as the only absorption mechanism. The right panel is showing the fitting of Eq. \ref{eq: parameterized} multiplied by Eq. \ref{eq: ff_od_param} to the spectra, assuming that the absorption is due to both SSA and FFA by the material in front of the shock. Both fits are also showing shaded lines uniformly drawn from the posteriors of the fitted parameters. \label{fig: ssa_ffa_compare}}}
\end{figure*}

We next model the first two radio spectra obtained by the VLA with the two-power-law model described above by fitting the spectra on $\Delta t = 201$ and $221$ days to Eq. \ref{eq: parameterized}. Since we do not observe the optically thin part of the spectrum on $\Delta t = 201$ days to high enough frequencies, we cannot determine the power law of the optically thin regime. Thus, in all further analysis, we will first fit the spectrum on $\Delta t = 221$ days and apply its fitted spectral slope to the fitting process of the spectrum on $\Delta t = 201$ days. For the spectrum on $\Delta t = 221$ days, the free parameters are the peak flux density, $F_{\rm \nu,a}$, its frequency, $\nu_{\rm a}$, and the optically thin power-law index, $\beta$ (we assume only SSA for now). We use \texttt{emcee} \citep{foreman_mackey_2013} to perform Markov chain Monte Carlo (MCMC) analysis to determine the posterior probability distributions of the parameters of the fitted model (and use flat priors). Based on the results of our fit (shown in the left panel of Fig. \ref{fig: ssa_ffa_compare}) we find that on $\Delta t = 221$ days $F_{\rm \nu,a} = 13.5 \pm 0.6$ mJy, $\nu_{\rm a} = 19.7 \pm 0.5$ GHz, and $\beta = 2.0 \pm 0.2$ (giving a reduced $\chi^2$ of $2.19$ for $15$ degrees of freedom, DOF, and a p-value of $0.005$). On $\Delta t = 201$ days the peak flux density is $17.8 \pm 0.5$ mJy, and its frequency is $22.8 \pm 0.3$ GHz (giving a reduced $\chi^2$ of $1.59$ for $16$ DOF, and a p-value of $0.06$).

This model fails to describe the rise of the optically thick regime and to match the flux density upper limit at $3$ GHz (see left panel of Figure \ref{fig: ssa_ffa_compare}). Furthermore, setting a significance level of $\alpha=0.05$, the p-value of the first epoch does not rule out that such a two-power-law model can describe the data. However, this is not the case for the second epoch which resulted in p-value$<0.05$. Since this model fails to reproduce the optically thick regime, we try to accommodate it by assuming that there is an additional absorption mechanism - external FFA by the material in front of the shock. 

We fit these two spectra with Eq. \ref{eq: parameterized} multiplied by $e^{-\tau_{\rm ff}}$ where the free parameters are the same as the ones we used above plus the FFA frequency $\nu_{\rm ff}$. We use \texttt{emcee} to perform MCMC analysis and determine the posterior probability distributions of the parameters of the fitted model (and use flat priors). Based on the results of our fit (shown in the right panel of Fig. \ref{fig: ssa_ffa_compare}) we find that on $\Delta t = 221$ days $F_{\rm \nu,a} = 13.6 \pm 0.7$ mJy, $\nu_{\rm a} = 17.9 \pm 0.6$ GHz, $\beta = 1.7 \pm 0.2$, and $\nu_{\rm ff} = 3.5 \pm 0.4$ GHz (giving a reduced $\chi^2$ of $0.58$ for $14$ DOF, and a p-value of $0.89$). On $\Delta t = 201$ days the peak flux density is $17.4 \pm 0.5$ mJy, its frequency is $21.5 \pm 0.4$ GHz, and the FFA frequency is $2.6 \pm 0.5$ GHz (giving a reduced $\chi^2$ of $0.69$ for $15$ DOF, and a p-value of $0.8$). 

Since the model of external FFA + SSA requires additional free parameter we also applied the Bayesian information criterion (BIC) for model selection. This criterion considers the model with the highest likelihood but penalizes models with more free parameters to avoid the issue of over-fitting. We find, for $\Delta t = 201$ days, that for the SSA-only model the BIC is $31.3$, and for the SSA + FFA model, the BIC is $19.2$ $\rm \left( \Delta BIC_{\rm 201 \, days} = 12.1 \right)$, and for $\Delta t = 221$ days, that for the SSA-only model the BIC is $41.6$, and for the SSA + FFA model, the BIC is $19.8$ $\rm \left( \Delta BIC_{\rm 221 \, days} = 21.8 \right)$. The difference between the BIC of the two models is relatively large ($\rm \Delta BIC>10$ for both epochs), thus, the SSA+FFA model is strongly favoured (see e.g., the discussion in \citealt{raftery1995bayesian}) Furthermore, the SSA+FFA model describes the rise of the optically thick regime better than the model without the additional FFA term for both epochs. It also matches the flux density upper limit at $3$ GHz despite the fact that we did not force it in our fitting procedure. We thus note that external FFA in addition to the internal SSA is a more probable scenario than just SSA. However, it is important to notice that while FFA is evident in the observed radio spectra, the peak flux density and frequency change only within up to $2\sigma$ between the two models.

Assuming a spherical and homogeneous synchrotron self-absorbed source, the physical parameters of the shock, such as its radius and magnetic field strength, are given by the peak flux density and its frequency (Eq. 11 and 12 in \citealt{chevalier_1998}; we assume here equipartition; $\epsilon_e = \epsilon_B = 0.1$, and an emission filling factor of $f=0.5$). We also assume that cooling effects, such as inverse-Compton (IC) and/or synchrotron cooling, are not in effect (however, synchrotron cooling is in effect; see relevant analysis in \S\ref{subsec: cooling}). In that case, the power-law index of the energy distribution, $p$, is given by the spectral index of the optically thin power law $\beta$, via $\beta=\left(p-1\right)/2$. The fitted parameters under the SSA + FFA model translate to a radius of $\left( 3.5 \pm 0.2 \right) \times 10^{16}$ cm, a magnetic field of $2.53 \pm 0.09$ G, and $p=4.4 \pm 0.4$, on $ \Delta t = 221$ days. On $\Delta t = 201$ days we infer $R = \left( 3.3 \pm 0.2 \right) \times 10^{16}$ cm, and $B = 2.98 \pm 0.07$ G.

Assuming constant expansion of the shock front, $v_{\rm sh} = R/\Delta t$, the shock velocity is $\rm \left( 1.9 \pm 0.1 \right) \times 10^4 \, \rm km \, s^{-1}$ on both epochs. Using Eq. \ref{eq: mass-loss} and a assuming wind velocity of $100 \, \rm km \, s^{-1}$ we find a mass-loss rate of $\dot{M} = \left( 1.9 \pm 0.3 \right) \times 10^{-4}$ and $\left( 1.6 \pm 0.1 \right) \times 10^{-3} \, \rm M_{\odot} \, yr^{-1}$ on $\Delta t = 201$ and $221$ days, respectively. The most likely progenitors of Type Ib SNe, He-stars, are expected to experience mass-loss at rates of $10^{-4} - 10^{-7} \, \rm M_{\odot} \, yr^{-1}$ \citep{smith_2014}. Thus, the mass-loss rates we infer from the radio spectral peaks are notably high.

\begin{figure}
\includegraphics[width=1\linewidth]{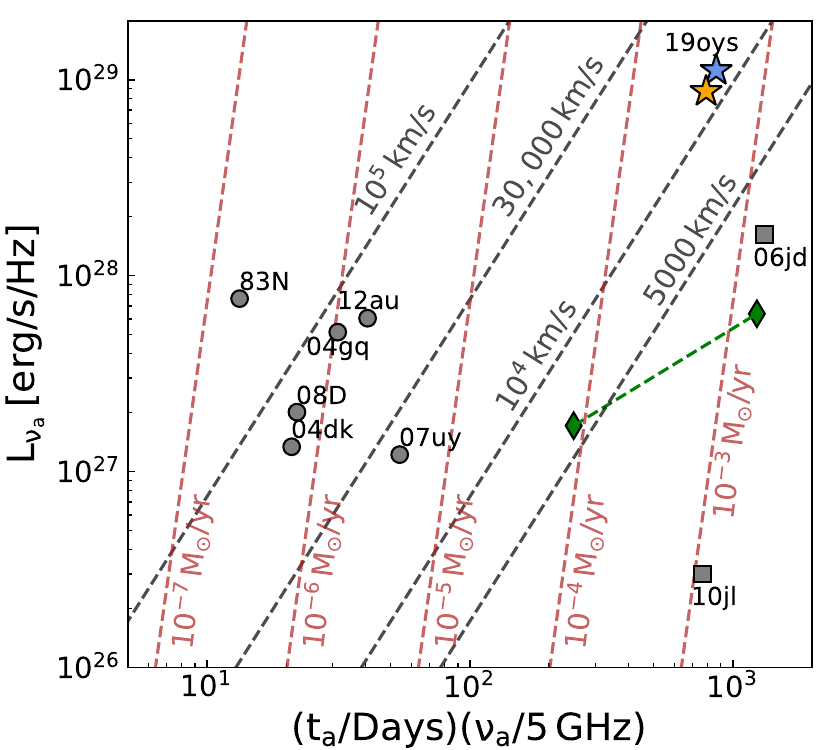}
\caption{\footnotesize{Chevalier diagram of SN\,2019oys (blue star marker for $\Delta t = 201$ days, and an orange star marker for $\Delta t = 221$ days) with other Type Ib (grey circles) and IIn (grey squares) SNe. Also plotted for reference (green diamonds) are the two peaks observed in the $15.7$ GHz light curve of SN\,2014C \citep{anderson_2017} which showed similar behaviour to SN\,2019oys in the optical. The diagram shows the peak spectral radio luminosity, $L_{\nu_a}$, against its frequency, $\nu_a$, and the time of the peak, $t_a$. Lines of equal shock velocity (black) and mass-loss rate (red; assuming wind velocity of 100 km/s) are plotted. We assumed here equipartition ($\epsilon_{\rm e}=\epsilon_{\rm B}=0.1$, $f=0.5$, and $p=3$ for the plotted lines. However, we note that we find $p>3$ from the spectral slope of the optically thin regime of SN 2019oys. Our assumption of $p=3$ is here to make the comparison between other CCSNe and SN 2019oys more straightforward as $p \simeq 3$ is usually observed for CCSNe. \label{fig: chev_diag}}}
\end{figure}

A useful way to compare the radio properties of SNe is by examining the phase space of peak spectral luminosity and its frequency times the observed peak time, for individual SNe. Lines of equal shock velocity and mass-loss rate are also plotted in this diagram (also called Chevalier's diagram; \citealt{chevalier_1998}). Fig. \ref{fig: chev_diag} is showing a Chevalier's diagram with the position of SN\,2019oys according to the radio spectral peak using the spectrum on $\Delta t = 201$ and $221$ days with other Type Ib and IIn SNe. In this diagram, we also plotted lines of equal velocity and mass-loss rate assuming the same assumption made above, and $p=3$. This assumption should be taken with care in the case of SN 2019oys as we observed optically thin spectral slopes that indicate $p>3$. However, since most CCSNe show $p \simeq 3$, we find this assumption more suitable for comparison with other CCSNe. As seen from this plot, SN\,2019oys is one the brightest SNe observed in radio wavelengths. SN\,2019oys is also slower than all Type Ib SNe. However, the velocity comparison to other Type Ib SNe is not straightforward as it is measured at a late time when usually the shock velocities from Type Ib SNe are measured at earlier times after the explosion. At late times, the shock accumulates mass which can slow it down significantly. We further discuss the temporal evolution of SN\,2019oys in \S\ref{subsec: temporal}.

\subsection{Electron cooling}
\label{subsec: cooling}

Typically, the power-law index of the electron energy distribution, $p$, is much smaller (typically $p=2.5-3.5$ for CCSNe, see e.g. \citealt{chevalier_1998}) than the value of $p=4.4 \pm 0.4$ we infer for $\Delta t = 221$ days. Such a high spectral slope can be an indication of cooling effects such as synchrotron and/or inverse-Compton (IC) cooling (see e.g. SN\,2012aw; \citealt{yadav_2014}, SN\,2013df; \citealt{kamble_2016}; SN\,2020oi; \citealt{horesh_2020}). The synchrotron cooling frequency is 
\begin{align}
\label{eq: sync_cool_freq}
    \nu_{\rm syn\_cool} = \frac{18 \pi m_e c q_e}{\sigma_T ^2 B^3 t^2},
\end{align}
where $m_e$ is the electron mass, $c$ is the speed of light, $q_e$ is the charge of the electron, and $\sigma_T$ is the Thomson cross-section. We find, based on Eq. \ref{eq: sync_cool_freq} and the values of magnetic fields we find in \S\ref{subsec: simple_sync}, that $\nu_{\rm syn\_cool} \leq 2$ GHz for both $\Delta t = 201$ and $221$ days, and therefore synchrotron cooling is important in our analysis. We do not test for IC cooling since its frequency is highly dependent on the bolometric luminosity of which we do not have any knowledge (see e.g. \citealt{chevalier_fransson_2006,horesh_2020}).

Taking synchrotron cooling into consideration, the spectral slope of the optically thin regime is now $\beta = p/2$, thus, we can derive $p=3.4 \pm 0.4$ for $\Delta t = 221$ days (remember that due to lack of coverage at high frequencies, we use this value of $p$ for both epochs). This translates, using the same analysis used in \S\ref{subsec: simple_sync}, to $R = \left( 2.4 \pm 0.1 \right) \times 10^{16}$ cm and $B = 1.78 \pm 0.04$ G on $\Delta t = 201$ days, and $R = \left( 2.6 \pm 0.2 \right) \times 10^{16}$ cm and $B = 1.51 \pm 0.05$ G on $\Delta t = 221$ days. Assuming constant expansion of the shock front, $v_{\rm sh} = R/\Delta t$, the shock velocity is $\left( 1.4 \pm 0.1 \right) \times 10^4 \, \rm km \, s^{-1}$ in both $\rm \Delta t = 201$ and $221$ days. From Eq. \ref{eq: mass-loss} (assuming $v_{\rm w} = 100 \, \rm km \, s^{-1}$) we derive a mass-loss rate of $\left( 6.7 \pm 0.3 \right)$ and $\left( 5.9 \pm 0.4 \right) \times 10^{-4} \, \rm M_{\odot} \, yr^{-1}$, for $\Delta t = 201$ and $221$ days, respectively. These are lower values of mass-loss rate (correspond to lower density) than inferred without taking electron cooling into consideration (see \S\ref{subsec: simple_sync}). The inferred electron temperature in the wind is $\left( 7 \pm 3 \right)$ and $\left( 3 \pm 1 \right) \times 10^4$ K, for $\Delta t = 201$ and $221$ days, respectively. A self-consistency check shows that, using the parameters inferred assuming synchrotron cooling, the cooling frequency is $\leq 1.3$ GHz for both epochs and therefore is important for the interpretation of the shock physical parameters.

\subsection{Temporal evolution}
\label{subsec: temporal}

Most of the frequencies observed with the VLA for SN\,2019oys do not have sufficient temporal coverage to analyze their temporal evolution. However our Ku-band observations with the VLA and AMI-LA ($15.5 \, \rm GHz$) spread from a few weeks to almost two years after optical discovery (right panel of Fig. \ref{fig: VLA_AMI}). In the following section, we will analyze the light curve of SN\,2019oys in the $15.5$ GHz band.

We examine here the temporal evolution of the SN by testing a two-power-law model such as the one introduced in Eq. \ref{eq: parameterized}
\begin{align}
    \label{eq: temporal_pl}
    F_{\nu} \left( \Delta t \right) = 1.582 \, F_{\nu,a} \left( \frac{\Delta t}{t_a} \right)^a \left( 1 - \exp\left[-\left( \frac{\Delta t}{t_a} \right)^{-(a+b)}\right] \right),
\end{align}
where $F_{\nu,a}$ is the peak flux at time $t_a$, and $a$ and $b$ are the power laws of the rising and declining regimes of the light curve, respectively.

\begin{figure}[!ht]
\includegraphics[width=1\linewidth]{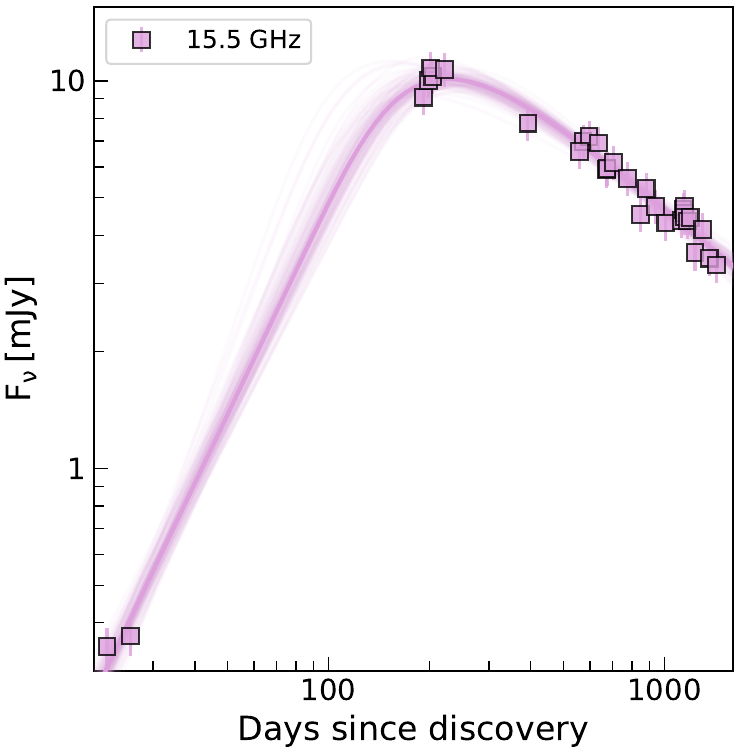}
\caption{\footnotesize{Modeling of the $15.5$ GHz light curve obtained with the VLA and AMI-LA. The light curve was fitted with a two-power-law model. Shaded lines are uniformly drawn parameters from the posterior probability distributions of the fitted parameters. \label{fig: tempoarl_ku}}}
\end{figure}

We fit the two-power-law model shown in Eq. \ref{eq: temporal_pl} to the $15.5$ GHz observations. The free parameters are the peak flux $F_{\nu,a}$, its time $t_a$, and the two power-law indices, $a$ and $b$. We use \texttt{emcee} to perform MCMC analysis to determine the posterior probability distributions of the parameters of the fitted model (and use flat priors). Based on the results of our fit (shown in Fig. \ref{fig: tempoarl_ku}) we find that $F_{\nu,a} = 9.8 \pm 0.4$ mJy, $t_a = 187 \pm 30$ days, $a = 1.8 ^{+0.2} _{-0.1}$, and $b = 0.71 \pm 0.07$. This results in $\chi^2 = 15.9$ for $26$ DOF (p-value$=0.97$). The observed decline rate of the $15.5$ GHz light curve is atypical for CCSNe. In the simple SN-CSM model, assuming interaction with a spherical and homogeneous CSM shell, with a density structure of $r^{-2}$, and free expansion, we expect a decline rate of $t^{-1}$ (for $p=3$, in our case of $p=3.4 \pm 0.4$ we expect $t^{-1.2 \pm 0.2}$; \citealt{chevalier_1998}). This discrepancy can be explained by an interaction with a more complex CSM structure at late times (see a detailed discussion on CSM inhomogeneities in \S\ref{sec: inhomogen}).

While the decline of the light curve is well constrained, the peak flux density, its time, and the rising power law are based on the two early data points. Since the AMI-LA beam is large ($\sim 30$ arcsec), it is possible that there is underlying emission from a diffusive background that contaminates our AMI-LA observations (and not the VLA observations thanks to its high resolution). Thus, while our late-time observations will not be affected too much by such contamination (as it is at most a few percent of the observed flux), it is possible that the two early observations are highly contaminated by a diffusive background. Therefore, we do not make use of these parameters to infer the physical properties of the shock and its surroundings.

\section{Inhomogeneities in the CSM}
\label{sec: inhomogen}

Broad radio spectral peaks have been observed previously in several SNe (e.g., SN\,2003L; \citealt{soderberg_2005}, PTF\,11qcj; \citealt{Bjornsson_2017}, Master OT J120451.50+265946.6; \citealt{Poonam_2019}). We also observe broad spectral peaks in SN\,2019oys on $\Delta t = 392$ and $557$ days. Such broad peaks cannot be explained using the standard SN-CSM interaction model with a homogeneous $r^{-2}$ CSM structure, and are usually attributed to inhomogeneities in the CSM structure caused by variations in the distribution of relativistic electrons and/or the magnetic field strength within the synchrotron source \citep{Bjornsson_2013, Bjornsson_2017}. Since the shock accelerates the electrons in the CSM and also enhances the magnetic fields, a non-homogeneous CSM structure can result in a varying magnetic field across the emitting region. Thus, a way of introducing a non-homogeneous CSM structure to the synchrotron source is by applying a varying magnetic field. 

\cite{Bjornsson_2013} has suggested that the magnetic field has a probability distribution of a power law, $P (B) \sim B^{- \alpha}$ between $B_{\rm 0}$ and $B_{\rm 1}$ (and zero elsewhere). Therefore, if the synchrotron spectrum for a given magnetic field $B$ is $F_\nu \left( R,B \right)$ (Eq. 1 in \citealt{chevalier_1998}); assuming no cooling effects), then the total spectrum observed for a non-homogeneous CSM is given by
\begin{align}
    \label{eq: inhomogen}
    F_{\rm \nu} = \int_{B_{\rm 0}} ^{B_{\rm 1}} P (B) F_{\rm \nu} \left( R,B \right) dB.
\end{align}
We next fit Eq. \ref{eq: inhomogen} to the spectra on days $392$ and $557$ after optical discovery where the free parameters are the radius of the emitting shell, $R$, and the parameters of the magnetic field probability distribution, $B_{\rm 0}$, $B_{\rm 1}$, and $\alpha$. In \S\ref{subsec: simple_sync} we found that the electron power-law index is $p=3.4 \pm 0.4$. Since we do not observe the optically thin regime to sufficiently high frequencies in the last two epochs we will use this value of $p$ in our analysis below.

\begin{figure}
\includegraphics[width=1\linewidth]{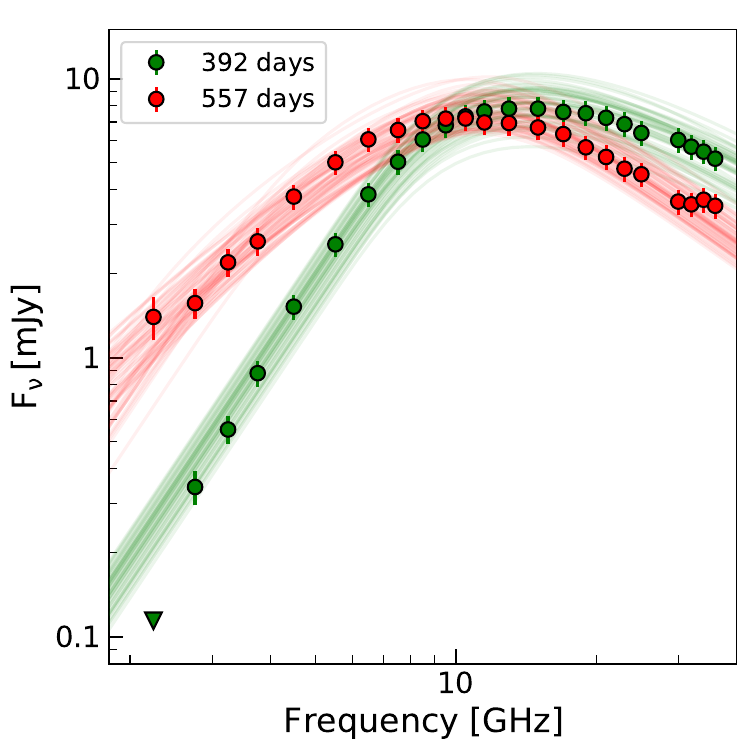}
\caption{\footnotesize{Modeling of the radio spectra on $\Delta t = 392$ and $557$ days according to the SN-CSM interaction model where inhomogeneities in the CSM are introduced through a magnetic field probability distribution (Eq. \ref{eq: inhomogen}). Shaded lines are uniformly drawn parameters from the posteriors of the fitted parameters. \label{fig: inhomogen}}}
\end{figure}

We use \texttt{emcee} \citep{foreman_mackey_2013} to perform MCMC analysis and determine the posterior probability distributions of the parameters of the fitted model (and use flat priors\footnote{Due to the long tail of the distributions of some of the parameters we fit to the base 10 logarithm of $B_1$ on $\Delta t = 392$ days, and of the radius and $B_0$ on $\Delta t = 557$ days (see Fig. \ref{fig: corner_392_bjor} and \ref{fig: corner_557_bjor} in appendix \ref{sec: corner_plots}).}). Based on the results of our fit for $\Delta t = 392$ (shown in Fig. \ref{fig: inhomogen}) we find that $R = \left( 3.4 \pm 0.2 \right) \times 10^{16} \, \rm cm$, $B_{\rm 0} = 0.82 \pm 0.03 \, \rm G$, $B_{\rm 1} \gtrsim 4 \, \rm G$, and $\alpha = 6.0 \pm 0.5$. Based on the results of our fit for $\Delta t = 557$ (shown in Fig. \ref{fig: inhomogen}) we find that $R \gtrsim 5 \times 10^{16} \, \rm cm$, $B_{\rm 0} \lesssim 0.45 \, \rm G$, $B_{\rm 1} =  1.2^{+0.5}_{-0.3} \, \rm G$, and $\alpha = 2.7^{+1.9}_{-1.2}$. Lines representing uniformly drawn parameters from the posteriors of the fitted parameters are plotted in Figure \ref{fig: inhomogen}. The shock velocity, assuming free expansion, is $\left(1.0 \pm 0.1 \right) \times 10^4 \, \rm km \, s^{-1}$ on $\Delta t = 392$ days, and $\gtrsim 10^4 \, \rm km \, s^{-1}$ on $\Delta t = 557$ days. This points toward a deceleration of the shock when compared to the velocities we infer on $\Delta t = 201$ and $221$ days (of $\sim 14,000$ km/s).

In this model, $B_0$ is the lowest magnetic field in the distribution and determines the transition frequency in the synchrotron spectrum from the optically thick regime to the spectral peak. $B_1$ is the highest magnetic field in the distribution and determines the transition from the spectral peak to the optically thin regime. Thus, in order to constrain these parameters one should observe the full optically thick, $F_{\rm \nu} \sim \nu^{5/2}$, to thin, $F_{\rm \nu} \sim \nu^{-(p-1)/2}$, synchrotron spectrum. As seen from our fitting process we got only a lower limit on $B_1$ for $\Delta t =392$ days. This is because, on $\Delta t =392$ days, we do not observe the transition from the spectral peak to the optically thin regime, and therefore cannot constrain $B_1$. On $\Delta t = 557$ days, however, we get only limits of $R$ and $B_0$. These limits are due to the fact that we do not observe the transition from the optically thick regime to the spectral peak due to the lack of sufficiently low frequencies. Furthermore, the posterior probability distributions of $B_1$ and $\alpha$ (see appendix \ref{sec: corner_plots} for corner plots) are broad due to a lack of spectral coverage. We also note that the probability distribution of the magnetic field varies significantly between the two epochs, both in the edge values ($B_0$ and $B_1$) and the power-law of the distribution, $\alpha$. This suggests significant variations in the CSM density structure between the two epochs.

A phenomenological model for describing broad spectral peaks in CCSNe was suggested by \cite{soderberg_2005}. The broadening of the spectral peak is artificially made by introducing a parameter $\zeta$ which flattens the self-absorbed synchrotron spectrum as seen in Eq. 3 in \cite{Bjornsson_2017}. In this model, $\zeta$ varies between $0$ to $1$, where $\zeta=1$ corresponds to a typical synchrotron spectrum with no flattening of the spectral peak. Implementation of this model was 
made by \cite{Bjornsson_2017} in the following selected SNe, SN\,2003L \citep{soderberg_2005} showed $\zeta=0.5$, SN\,2003bg \citep{soderberg_2006} showed $\zeta=0.6$, and the most extreme spectrum flattening was observed for PTF\,11qcj \citep{corsi_2014} with $\zeta=0.2$. We find, for SN\,2019oys, that on $\Delta t = 392$ days $\zeta = 0.65 \pm 0.07$, and on $\Delta t = 557$ days $\zeta = 0.46 \pm 0.04$. We note here that since the spectrum on $\Delta t = 392$ days does not exhibit the transition from the peak to the optically thin regime, the inferred broadening parameter might decrease if the spectral peak is even broader than observed.

Overall, SN\,2019oys is showing one of the broadest radio spectral peaks observed so far, which may be an indication of it being surrounded by an extremely non-homogeneous CSM structure, even compared to other notable SNe. We do note that if the peak broadness is a measure of the CSM in-homogeneity, the shock traveling in the CSM first encounters a rather homogeneous CSM structure, as seen by the radio spectra on $\Delta t = 201$ and $221$ days, and only later, the CSM structure becomes less homogeneous. This is somewhat puzzling since distant CSM shells were deposited earlier in the star evolution (prior to the explosion) and had more time to become more homogeneous. It is possible that the CSM was formed in two (or more) distinct phases of mass loss resulting in two different types of CSM structures, one observed on $\Delta t = 201$ and $221$ days, and the other on $\Delta t = 392$ and $557$ days.

\section{Conclusions}
\label{sec: conclusions}

In this work, we have presented multi-epoch, broadband, radio observations of the peculiar case of SN\,2019oys. SN\,2019oys was first classified as a Type Ib SN but after about $\sim 100$ days, it showed optical rebrightening and a spectral makeover to a spectrum mainly dominated by CSM interaction and a plethora of narrow high-ionization lines, including coronal lines. About $150$ days after the rebrightening the optical light curve began to slowly fade, and since then continue to decline.

Our broadband radio observations presented here, taken with the VLA and AMI-LA, show a bright radio source at late times, which is in agreement with the strong CSM interaction seen in the optical. Analyzing the radio spectra on $\Delta t = 201$ and $221$ days has shown that the radio emission from a synchrotron self-absorbed source, is accompanied by a FFA from the material in front of the shock. Initial analysis, not taking electron cooling into consideration, of the radio spectral peaks suggested magnetic fields of $\rm 1 - 2 \, G$ at a relatively late time after the explosion. In addition, the power-law index of the electron energy distribution was extremely high ($p=4.4 \pm 0.4$) for the spectrum obtained on $\Delta t = 221$ days. Such a steep power-law index and high magnetic fields point toward electron cooling (either by synchrotron and/or IC cooling). We find that the synchrotron cooling frequency is lower than all of our observed frequencies on $\Delta t = 201$ and $221$ days, thus, making it important for our analysis. We do not test for IC cooling as its frequency is highly dependent on the bolometric luminosity, which we do not know.

Taking synchrotron cooling into consideration we find that the power-law index of the electron energy distribution is $3.4 \pm 0.4$ on $\Delta t = 221$ days. Since we lack the coverage of high frequencies in the optically thin regime on $\Delta t = 201$ days, we cannot constrain the power-law index, $p$, at that epoch, and we assume that it does not change between epochs. Using the spectral peak, and assuming that the CSM was deposited by constant mass-loss in steady winds from the progenitor massive star, we find that the mass-loss rate, as observed in $\Delta t = 201$ and $221$ days, is $\left( 5.9 \pm 0.4\right)$ and $\left( 6.7 \pm 0.3\right)$ $\times 10^{-4} \, \rm M_{\odot} \, yr^{-1}$ (assuming wind velocity of $100 \, \rm km \, s^{-1}$). This translates into a density of $\rm \left( 2.6 \pm 0.5 \right) \times 10^{5} \, cm^{-3}$ at a distance of $\rm \left( 2.6 \pm 0.2 \right) \times 10^{16} \, cm$ which agrees with the CSM density of the region responsible for the narrow, high-ionization lines, seen in optical spectra \citep{sollerman_2020}.

The temporal evolution of SN\,2019oys is slowed when compared to other Type Ib SNe. The $15.5$ GHz light curve shows a slow decline of $t^{-0.71 \pm 0.07}$, which is also slower than the $t^{-1}$ expected in the scenario of a shock traveling with a constant velocity in a spherical, $r^{-2}$, density structure \citep{chevalier_1998}. Furthermore, the analysis above suggests a decelerating shock velocity of $\left( 1.4 \pm 0.1 \right) \times 10^4 \, \rm km\,s^{-1}$ on $\Delta t = 201$ and $221$ days, and $\left( 1.0 \pm 0.1 \right) \times 10^4 \, \rm km\,s^{-1}$ on $\Delta t = 392$ days. These values are lower than the velocities observed in radio for most Type Ib SNe. However, typically, velocities of Type Ib SNe are measured early on (within weeks to a few months after the explosion) and the velocity of SN\,2019oys was measured about 7 months after optical discovery. Furthermore, Type IIn SNe are typically slowed down by the dense CSM surrounding the progenitor. Thus, it is possible that this low velocity is due to mass accumulation by the shock traveling in a dense CSM (which is also supported by the observed shock deceleration).

SN\,2019oys is also showing late-time radio spectra with unusually broad spectral peaks. These spectral peaks can be interpreted as a synchrotron emitting source from a non-homogeneous CSM structure. We fitted a model describing the CSM in-homogeneities as arising from variations in the distribution of the relativistic electrons and/or the magnetic field strength, by introducing a power-law distribution for the magnetic field strength as presented in \cite{Bjornsson_2013, Bjornsson_2017}. This model manages to describe the data, however, due to the lack of spectral coverage we can only put upper limits on some of the model parameters. Comparing the spectral peak broadness of SN\,2019oys to other notable SNe shows that the CSM structure around the SN is highly non-homogeneous, however, broader peaks have been observed in several cases (e.g. PTF\,11qcj; \citealt{corsi_2014}).

Overall, SN\,2019oys is showing strong CSM interaction at late times, both in optical and radio. It shows significant signs of FFA absorption in radio, implying high density at a large distance, and, a non-homogeneous CSM structure. The evolution of the radio spectrum of SN\,2019oys raises interesting questions on the formation of the CSM structure encompassing its progenitor. For example, it is somewhat surprising that the outer regions (observed in the late-time radio spectra) are non-homogeneous, while the inner CSM (observed in the early radio spectra) points toward a rather homogeneous CSM shell (without a broad radio spectral peak). Such variations in the CSM structure over relatively short time scales prior to the SN explosion can be an indication of several, distinct mass-loss episodes at the end of the massive star life. This is somewhat reminiscence of the optical signatures of extreme mass-loss episodes, about a year prior to the explosion, that is sometimes seen in recombination emission lines via the technique of flash spectroscopy - rapid spectroscopic observations of SNe, shortly (hours) after they explode \citep{gal-yam_2014}.

SN\,2019oys is demonstrating the importance of late-time observations of SNe in optical, and especially in radio. Late-time radio observations play a key role in revealing the mass-loss history of massive stars in different epochs of their evolution.

\begin{acknowledgements}
A.H. is grateful for the support by the Israel Science Foundation (ISF grant 1679/23) and by the the United States-Israel Binational Science Foundation (BSF grant 2020203) We acknowledge the staff who operate and run the AMI-LA telescope at Lord's Bridge, Cambridge, for the AMI-LA radio data. AMI is supported by the Universities of Cambridge and Oxford, and by the European Research Council under grant ERC-2012-StG-307215 LODESTONE. We thank the National Radio Astronomy Observatory (NRAO) for carrying out the Karl G. Jansky Very Large Array (VLA) observations. D.R.A.W was supported by the Oxford Centre for Astrophysical Surveys, which is funded through generous support from the Hintze Family Charitable Foundation. S. Schulze is partially supported by LBNL Subcontract NO. 7707915 and by the G.R.E.A.T. research environment, funded by {\em Vetenskapsr\aa det},  the Swedish Research Council, project number 2016-06012. This research has made use of the CIRADA cutout service at URL cutouts.cirada.ca, operated by the Canadian Initiative for Radio Astronomy Data Analysis (CIRADA). CIRADA is funded by a grant from the Canada Foundation for Innovation 2017 Innovation Fund (Project 35999), as well as by the Provinces of Ontario, British Columbia, Alberta, Manitoba and Quebec, in collaboration with the National Research Council of Canada, the US National Radio Astronomy Observatory and Australia’s Commonwealth Scientific and Industrial Research Organisation.

We have made extensive use of the data reduction software {\sc casa} \citep{2007ASPC..376..127M}, {\sc reduce$\_$dc} \citep{perrott_2013}, and the {\sc python} {\sc emcee} \citep{foreman_mackey_2013} package.
        
\end{acknowledgements}

\bibliographystyle{aa}
\bibliography{main.bib}

\begin{thebibliography}{37}
\expandafter\ifx\csname natexlab\endcsname\relax\def\natexlab#1{#1}\fi

\bibitem[{{Anderson} {et~al.}(2017){Anderson}, {Horesh}, {Mooley}, {Rushton},
  {Fender}, {Staley}, {Argo}, {Beswick}, {Hancock}, {P{\'e}rez-Torres},
  {Perrott}, {Plotkin}, {Pretorius}, {Rumsey}, \&
  {Titterington}}]{anderson_2017}
{Anderson}, G.~E., {Horesh}, A., {Mooley}, K.~P., {et~al.} 2017, MNRAS, 466,
  3648

\bibitem[{{Bellm} {et~al.}(2019){Bellm}, {Kulkarni}, {Barlow}, {Feindt},
  {Graham}, {Goobar}, {Kupfer}, {Ngeow}, {Nugent}, {Ofek}, {Prince}, {Riddle},
  {Walters}, \& {Ye}}]{Bellm_2019}
{Bellm}, E.~C., {Kulkarni}, S.~R., {Barlow}, T., {et~al.} 2019, \pasp, 131,
  068003

\bibitem[{{Bj{\"o}rnsson}(2013)}]{Bjornsson_2013}
{Bj{\"o}rnsson}, C.~I. 2013, \apj, 769, 65

\bibitem[{{Bj{\"o}rnsson} \& {Keshavarzi}(2017)}]{Bjornsson_2017}
{Bj{\"o}rnsson}, C.~I. \& {Keshavarzi}, S.~T. 2017, \apj, 841, 12

\bibitem[{{Chandra} {et~al.}(2019){Chandra}, {Nayana}, {Bj{\"o}rnsson},
  {Taddia}, {Lundqvist}, {Ray}, \& {Shappee}}]{Poonam_2019}
{Chandra}, P., {Nayana}, A.~J., {Bj{\"o}rnsson}, C.~I., {et~al.} 2019, \apj,
  877, 79

\bibitem[{Chevalier(1981)}]{chevalier_1981}
Chevalier, R.~A. 1981, \apj, 251, 259

\bibitem[{Chevalier(1982)}]{chevalier_radio_1982}
Chevalier, R.~A. 1982, \apj, 259, 302

\bibitem[{Chevalier(1998)}]{chevalier_1998}
Chevalier, R.~A. 1998, \apj, 499, 810

\bibitem[{Chevalier \& Fransson(2006)}]{chevalier_fransson_2006}
Chevalier, R.~A. \& Fransson, C. 2006, \apj, 651, 381

\bibitem[{{Corsi} {et~al.}(2014){Corsi}, {Ofek}, {Gal-Yam}, {Frail},
  {Kulkarni}, {Fox}, {Kasliwal}, {Sullivan}, {Horesh}, {Carpenter}, {Maguire},
  {Arcavi}, {Cenko}, {Cao}, {Mooley}, {Pan}, {Sesar}, {Sternberg}, {Xu},
  {Bersier}, {James}, {Bloom}, \& {Nugent}}]{corsi_2014}
{Corsi}, A., {Ofek}, E.~O., {Gal-Yam}, A., {et~al.} 2014, \apj, 782, 42

\bibitem[{{DeMarchi} {et~al.}(2022){DeMarchi}, {Margutti}, {Dittman},
  {Brunthaler}, {Milisavljevic}, {Bietenholz}, {Stauffer}, {Brethauer},
  {Coppejans}, {Auchettl}, {Alexander}, {Kilpatrick}, {Bright}, {Kelley},
  {Stroh}, \& {Jacobson-Gal{\'a}n}}]{demarchi_2022}
{DeMarchi}, L., {Margutti}, R., {Dittman}, J., {et~al.} 2022, \apj, 938, 84

\bibitem[{{Filippenko}(1997)}]{Filippenko_1997}
{Filippenko}, A.~V. 1997, \araa, 35, 309

\bibitem[{{Foreman-Mackey} {et~al.}(2013){Foreman-Mackey}, {Conley},
  {Meierjurgen Farr}, {Hogg}, {Lang}, {Marshall}, {Price-Whelan}, {Sanders}, \&
  {Zuntz}}]{foreman_mackey_2013}
{Foreman-Mackey}, D., {Conley}, A., {Meierjurgen Farr}, W., {et~al.} 2013,
  {emcee: The MCMC Hammer}

\bibitem[{{Gal-Yam}(2017)}]{SN_handbook}
{Gal-Yam}, A. 2017, in Handbook of Supernovae, ed. A.~W. {Alsabti} \&
  P.~{Murdin}, 195

\bibitem[{Gal-Yam {et~al.}(2014)Gal-Yam, Arcavi, Ofek, Ben-Ami, Cenko,
  Kasliwal, Cao, Yaron, Tal, Silverman, Horesh, De~Cia, Taddia, Sollerman,
  Perley, Vreeswijk, Kulkarni, Nugent, Filippenko, \& Wheeler}]{gal-yam_2014}
Gal-Yam, A., Arcavi, I., Ofek, E.~O., {et~al.} 2014, \nat, 509, 471

\bibitem[{{Graham} {et~al.}(2019){Graham}, {Kulkarni}, {Bellm}, {Adams},
  {Barbarino}, {Blagorodnova}, {Bodewits}, {Bolin}, {Brady}, {Cenko}, {Chang},
  {Coughlin}, {De}, {Eadie}, {Farnham}, {Feindt}, {Franckowiak}, {Fremling},
  {Gezari}, {Ghosh}, {Goldstein}, {Golkhou}, {Goobar}, {Ho}, {Huppenkothen},
  {Ivezi{\'c}}, {Jones}, {Juric}, {Kaplan}, {Kasliwal}, {Kelley}, {Kupfer},
  {Lee}, {Lin}, {Lunnan}, {Mahabal}, {Miller}, {Ngeow}, {Nugent}, {Ofek},
  {Prince}, {Rauch}, {van Roestel}, {Schulze}, {Singer}, {Sollerman}, {Taddia},
  {Yan}, {Ye}, {Yu}, {Barlow}, {Bauer}, {Beck}, {Belicki}, {Biswas}, {Brinnel},
  {Brooke}, {Bue}, {Bulla}, {Burruss}, {Connolly}, {Cromer}, {Cunningham},
  {Dekany}, {Delacroix}, {Desai}, {Duev}, {Feeney}, {Flynn}, {Frederick},
  {Gal-Yam}, {Giomi}, {Groom}, {Hacopians}, {Hale}, {Helou}, {Henning},
  {Hover}, {Hillenbrand}, {Howell}, {Hung}, {Imel}, {Ip}, {Jackson}, {Kaspi},
  {Kaye}, {Kowalski}, {Kramer}, {Kuhn}, {Landry}, {Laher}, {Mao}, {Masci},
  {Monkewitz}, {Murphy}, {Nordin}, {Patterson}, {Penprase}, {Porter},
  {Rebbapragada}, {Reiley}, {Riddle}, {Rigault}, {Rodriguez}, {Rusholme}, {van
  Santen}, {Shupe}, {Smith}, {Soumagnac}, {Stein}, {Surace}, {Szkody}, {Terek},
  {Van Sistine}, {van Velzen}, {Vestrand}, {Walters}, {Ward}, {Zhang}, \&
  {Zolkower}}]{Graham_2019}
{Graham}, M.~J., {Kulkarni}, S.~R., {Bellm}, E.~C., {et~al.} 2019, \pasp, 131,
  078001

\bibitem[{Hickish {et~al.}(2018)Hickish, Razavi-Ghods, Perrott, Titterington,
  Carey, Scott, Grainge, Scaife, Alexander, Saunders, Crofts, Javid, Rumsey,
  Jin, Ely, Shaw, Northrop, Pooley, D'Alessandro, Doherty, \&
  Willatt}]{hickish_2018}
Hickish, J., Razavi-Ghods, N., Perrott, Y.~C., {et~al.} 2018, \mnras, 475, 5677

\bibitem[{{Horesh} {et~al.}(2012){Horesh}, {Kulkarni}, {Fox}, {Carpenter},
  {Kasliwal}, {Ofek}, {Quimby}, {Gal-Yam}, {Cenko}, {de Bruyn}, {Kamble},
  {Wijers}, {van der Horst}, {Kouveliotou}, {Podsiadlowski}, {Sullivan},
  {Maguire}, {Howell}, {Nugent}, {Gehrels}, {Law}, {Poznanski}, \&
  {Shara}}]{horesh_2012}
{Horesh}, A., {Kulkarni}, S.~R., {Fox}, D.~B., {et~al.} 2012, \apj, 746, 21

\bibitem[{{Horesh} {et~al.}(2020){Horesh}, {Sfaradi}, {Ergon}, {Barbarino},
  {Sollerman}, {Moldon}, {Dobie}, {Schulze}, {P{\'e}rez-Torres}, {Williams},
  {Fremling}, {Gal-Yam}, {Kulkarni}, {O'Brien}, {Lundqvist}, {Murphy},
  {Fender}, {Anand}, {Belicki}, {Bellm}, {Coughlin}, {De}, {Golkhou}, {Graham},
  {Green}, {Hankins}, {Kasliwal}, {Kupfer}, {Laher}, {Masci}, {Miller},
  {Neill}, {Ofek}, {Perrott}, {Porter}, {Reiley}, {Rigault}, {Rodriguez},
  {Rusholme}, {Shupe}, \& {Titterington}}]{horesh_2020}
{Horesh}, A., {Sfaradi}, I., {Ergon}, M., {et~al.} 2020, \apj, 903, 132

\bibitem[{Kamble {et~al.}(2016)Kamble, Margutti, Soderberg, Chakraborti,
  Fransson, Chevalier, Powell, Milisavljevic, Parrent, \&
  Bietenholz}]{kamble_2016}
Kamble, A., Margutti, R., Soderberg, A.~M., {et~al.} 2016, \apj, 818, 111

\bibitem[{{Lacy} {et~al.}(2020){Lacy}, {Baum}, {Chandler}, {Chatterjee},
  {Clarke}, {Deustua}, {English}, {Farnes}, {Gaensler}, {Gugliucci},
  {Hallinan}, {Kent}, {Kimball}, {Law}, {Lazio}, {Marvil}, {Mao}, {Medlin},
  {Mooley}, {Murphy}, {Myers}, {Osten}, {Richards}, {Rosolowsky}, {Rudnick},
  {Schinzel}, {Sivakoff}, {Sjouwerman}, {Taylor}, {White}, {Wrobel},
  {Andernach}, {Beasley}, {Berger}, {Bhatnager}, {Birkinshaw}, {Bower},
  {Brandt}, {Brown}, {Burke-Spolaor}, {Butler}, {Comerford}, {Demorest}, {Fu},
  {Giacintucci}, {Golap}, {G{\"u}th}, {Hales}, {Hiriart}, {Hodge}, {Horesh},
  {Ivezi{\'c}}, {Jarvis}, {Kamble}, {Kassim}, {Liu}, {Loinard}, {Lyons},
  {Masters}, {Mezcua}, {Moellenbrock}, {Mroczkowski}, {Nyland}, {O'Dea},
  {O'Sullivan}, {Peters}, {Radford}, {Rao}, {Robnett}, {Salcido}, {Shen},
  {Sobotka}, {Witz}, {Vaccari}, {van Weeren}, {Vargas}, {Williams}, \&
  {Yoon}}]{vlass}
{Lacy}, M., {Baum}, S.~A., {Chandler}, C.~J., {et~al.} 2020, \pasp, 132, 035001

\bibitem[{{Langer}(2012)}]{Langer_2012}
{Langer}, N. 2012, \araa, 50, 107

\bibitem[{{McMullin} {et~al.}(2007){McMullin}, {Waters}, {Schiebel}, {Young},
  \& {Golap}}]{2007ASPC..376..127M}
{McMullin}, J.~P., {Waters}, B., {Schiebel}, D., {Young}, W., \& {Golap}, K.
  2007, Astronomical Society of the Pacific Conference Series, Vol. 376, {CASA
  Architecture and Applications}, 127

\bibitem[{{Nayana} {et~al.}(2018){Nayana}, {Chandra}, \& {Ray}}]{nayana_2018}
{Nayana}, A.~J., {Chandra}, P., \& {Ray}, A.~K. 2018, \apj, 863, 163

\bibitem[{Perrott {et~al.}(2013)Perrott, Scaife, Green, Davies, Franzen,
  Grainge, Hobson, Hurley-Walker, Lasenby, Olamaie, Pooley,
  Rodríguez-Gonzálvez, Rumsey, Saunders, Schammel, Scott, Shimwell,
  Titterington, Waldram, \& {AMI Consortium}}]{perrott_2013}
Perrott, Y.~C., Scaife, A. M.~M., Green, D.~A., {et~al.} 2013, \mnras, 429,
  3330

\bibitem[{Raftery(1995)}]{raftery1995bayesian}
Raftery, A.~E. 1995, Sociological methodology, 111

\bibitem[{{Ruiz-Carmona} {et~al.}(2022){Ruiz-Carmona}, {Sfaradi}, \&
  {Horesh}}]{SN2016x_2022}
{Ruiz-Carmona}, R., {Sfaradi}, I., \& {Horesh}, A. 2022, \aap, 666, A82

\bibitem[{Smith(2014)}]{smith_2014}
Smith, N. 2014, ARA\&A, 52, 487

\bibitem[{{Soderberg} {et~al.}(2006){Soderberg}, {Chevalier}, {Kulkarni}, \&
  {Frail}}]{soderberg_2006}
{Soderberg}, A.~M., {Chevalier}, R.~A., {Kulkarni}, S.~R., \& {Frail}, D.~A.
  2006, \apj, 651, 1005

\bibitem[{{Soderberg} {et~al.}(2005){Soderberg}, {Kulkarni}, {Berger},
  {Chevalier}, {Frail}, {Fox}, \& {Walker}}]{soderberg_2005}
{Soderberg}, A.~M., {Kulkarni}, S.~R., {Berger}, E., {et~al.} 2005, \apj, 621,
  908

\bibitem[{{Sollerman} {et~al.}(2020){Sollerman}, {Fransson}, {Barbarino},
  {Fremling}, {Horesh}, {Kool}, {Schulze}, {Sfaradi}, {Yang}, {Bellm},
  {Burruss}, {Cunningham}, {De}, {Drake}, {Golkhou}, {Green}, {Kasliwal},
  {Kulkarni}, {Kupfer}, {Laher}, {Masci}, {Rodriguez}, {Rusholme}, {Williams},
  {Yan}, \& {Zolkower}}]{sollerman_2020}
{Sollerman}, J., {Fransson}, C., {Barbarino}, C., {et~al.} 2020, \aap, 643, A79

\bibitem[{{Weiler} {et~al.}(2002){Weiler}, {Panagia}, {Montes}, \&
  {Sramek}}]{weiler_2002}
{Weiler}, K.~W., {Panagia}, N., {Montes}, M.~J., \& {Sramek}, R.~A. 2002,
  \araa, 40, 387

\bibitem[{{Weiler} {et~al.}(1991){Weiler}, {van Dyk}, {Panagia}, {Sramek}, \&
  {Discenna}}]{weiler_1991}
{Weiler}, K.~W., {van Dyk}, S.~D., {Panagia}, N., {Sramek}, R.~A., \&
  {Discenna}, J.~L. 1991, ApJ, 380, 161

\bibitem[{{Weiler} {et~al.}(2007){Weiler}, {Williams}, {Panagia}, {Stockdale},
  {Kelley}, {Sramek}, {Van Dyk}, \& {Marcaide}}]{weiler_2007}
{Weiler}, K.~W., {Williams}, C.~L., {Panagia}, N., {et~al.} 2007, ApJ, 671,
  1959

\bibitem[{Yadav {et~al.}(2014)Yadav, Ray, Chakraborti, Stockdale, Chandra,
  Smith, Roy, Bose, Dwarkadas, Sutaria, \& Pooley}]{yadav_2014}
Yadav, N., Ray, A., Chakraborti, S., {et~al.} 2014, \apj, 782, 30

\bibitem[{{Zenati} {et~al.}(2022){Zenati}, {Wang}, {Bobrick}, {DeMarchi},
  {Glanz}, {Rozner}, {Rest}, {Metzger}, {Margutti}, {Gomez}, {Smith}, {Toonen},
  {Bright}, {Norman}, {Foley}, {Gagliano}, {Krolik}, {Smartt}, {Villar},
  {Narayan}, {Fox}, {Auchettl}, {Brethauer}, {Clocchiatti}, {Coelln},
  {Coppejans}, {Dimitriadis}, {Doroszmai}, {Drout}, {Jacobson-Galan}, {Gao},
  {Ridden-Harper}, {Kilpatrick}, {Laskar}, {Matthews}, {Rest}, {Smith},
  {McKenzie Stauffer}, {Stroh}, {Strolger}, {Terreran}, {Pierel}, \&
  {Piro}}]{SN2019tsf}
{Zenati}, Y., {Wang}, Q., {Bobrick}, A., {et~al.} 2022, arXiv e-prints,
  arXiv:2207.07146

\bibitem[{Zwart {et~al.}(2008)Zwart, Barker, Biddulph, Bly, Boysen, Brown,
  Clementson, Crofts, Culverhouse, Czeres, Dace, Davies, D'Alessandro, Doherty,
  Duggan, Ely, Felvus, Feroz, Flynn, Franzen, Geisbüsch, Génova-Santos,
  Grainge, Grainger, Hammett, Hills, Hobson, Holler, Hurley-Walker, Jilley,
  Jones, Kaneko, Kneissl, Lancaster, Lasenby, Marshall, Newton, Norris,
  Northrop, Odell, Petencin, Pober, Pooley, Pospieszalski, Quy,
  Rodríguez-Gonzálvez, Saunders, Scaife, Schofield, Scott, Shaw, Shimwell,
  Smith, Taylor, Titterington, Velić, Waldram, West, Wood, Yassin, \& {AMI
  Consortium}}]{zwart_2008}
Zwart, J. T.~L., Barker, R.~W., Biddulph, P., {et~al.} 2008, \mnras, 391, 1545

\end{thebibliography}

\onecolumn
\begin{appendix}

\section{Tables}
\label{sec: tables}

\begin{longtable}{cccccc}
\caption{SN\,2019oys - radio observations\label{tab:Radio_Observations}}\\
\hline\hline
$\Delta t$ & $\nu$ & Band & $F_{\nu}$ & RMS & Telescope \\
$\rm \left[Day\right]$ & $\rm \left[GHz\right]$ &  & $\rm \left[mJy\right]$ & $\rm \left[mJy\right]$ &  \\
\hline
\endfirsthead
\caption{continued.}\\
\hline\hline
$\Delta t$ & $\nu$ & Band & $F_{\nu}$ & RMS & Telescope \\
$\rm \left[Day\right]$ & $\rm \left[GHz\right]$ &  & $\rm \left[mJy\right]$ & $\rm \left[mJy\right]$ &  \\
\hline
\endhead
\hline
\endfoot
$22$ & $15.5$ & Ku & $0.35 \pm 0.05$ & $0.04$ & AMI-LA \\ [0.1ex]
$26$ & $15.5$ & Ku & $0.37 \pm 0.06$ & $0.04$ & AMI-LA \\ [0.1ex]
$191$ & $15.5$ & Ku & $9.1 \pm 0.9$ & $0.06$ & AMI-LA \\ [0.1ex]
$198$ & $15.5$ & Ku & $10 \pm 1$ & $0.06$ & AMI-LA \\ [0.1ex]
$202$ & $3$ & S & $<0.09$ & $0.03$ & VLA:C \\ [0.1ex]
$202$ & $4.5$ & C & $0.39 \pm 0.04$ & $0.02$ & VLA:C \\ [0.1ex]
$202$ & $5.5$ & C & $0.70 \pm 0.07$ & $0.02$ & VLA:C \\ [0.1ex]
$201$ & $6.5$ & C & $1.2 \pm 0.1$ & $0.02$ & VLA:C \\ [0.1ex]
$201$ & $7.5$ & C & $1.8 \pm 0.2$ & $0.02$ & VLA:C \\ [0.1ex]
$201$ & $8.5$ & X & $2.6 \pm 0.3$ & $0.02$ & VLA:C \\ [0.1ex]
$201$ & $9.5$ & X & $3.7 \pm 0.4$ & $0.02$ & VLA:C \\ [0.1ex]
$201$ & $10.5$ & X & $4.8 \pm 0.5$ & $0.02$ & VLA:C \\ [0.1ex]
$201$ & $11.5$ & X & $6.0 \pm 0.6$ & $0.02$ & VLA:C \\ [0.1ex]
$201$ & $13$ & Ku & $8.1 \pm 0.8$ & $0.02$ & VLA:C \\ [0.1ex]
$201$ & $15$ & Ku & $10 \pm 1$ & $0.02$ & VLA:C \\ [0.1ex]
$201$ & $17$ & Ku & $13 \pm 1$ & $0.03$ & VLA:C \\ [0.1ex]
$201$ & $19$ & K & $15 \pm 1$ & $0.02$ & VLA:C \\ [0.1ex]
$201$ & $21$ & K & $15 \pm 2$ & $0.02$ & VLA:C \\ [0.1ex]
$201$ & $23$ & K & $16 \pm 2$ & $0.02$ & VLA:C \\ [0.1ex]
$201$ & $25$ & K & $16 \pm 2$ & $0.02$ & VLA:C \\ [0.1ex]
$201$ & $30$ & Ka & $14 \pm 1$ & $0.02$ & VLA:C \\ [0.1ex]
$201$ & $32$ & Ka & $14 \pm 1$ & $0.02$ & VLA:C \\ [0.1ex]
$201$ & $34$ & Ka & $13 \pm 1$ & $0.03$ & VLA:C \\ [0.1ex]
$201$ & $36$ & Ka & $12 \pm 1$ & $0.03$ & VLA:C \\ [0.1ex]
$204$ & $15.5$ & Ku & $10 \pm 1$ & $0.05$ & AMI-LA \\ [0.1ex]
$221$ & $3$ & S & $<0.1$ & $0.03$ & VLA:C \\ [0.1ex]
$221$ & $4.5$ & C & $0.37 \pm 0.04$ & $0.02$ & VLA:C \\ [0.1ex]
$221$ & $5.5$ & C & $0.71 \pm 0.08$ & $0.02$ & VLA:C \\ [0.1ex]
$221$ & $6.5$ & C & $1.4 \pm 0.2$ & $0.02$ & VLA:C \\ [0.1ex]
$221$ & $7.5$ & C & $2.1 \pm 0.2$ & $0.02$ & VLA:C \\ [0.1ex]
$221$ & $8.5$ & X & $3.1 \pm 0.3$ & $0.02$ & VLA:C \\ [0.1ex]
$221$ & $9.5$ & X & $4.2 \pm 0.4$ & $0.02$ & VLA:C \\ [0.1ex]
$221$ & $10.5$ & X & $5.4 \pm 0.6$ & $0.02$ & VLA:C \\ [0.1ex]
$221$ & $11.5$ & X & $6.5 \pm 0.7$ & $0.02$ & VLA:C \\ [0.1ex]
$221$ & $13$ & Ku & $8.6 \pm 0.9$ & $0.03$ & VLA:C \\ [0.1ex]
$221$ & $15$ & Ku & $10 \pm 1$ & $0.03$ & VLA:C \\ [0.1ex]
$221$ & $17$ & Ku & $12 \pm 1$ & $0.04$ & VLA:C \\ [0.1ex]
$221$ & $19$ & K & $13 \pm 1$ & $0.02$ & VLA:C \\ [0.1ex]
$221$ & $21$ & K & $13 \pm 1$ & $0.03$ & VLA:C \\ [0.1ex]
$221$ & $23$ & K & $13 \pm 1$ & $0.04$ & VLA:C \\ [0.1ex]
$221$ & $25$ & K & $12 \pm 1$ & $0.05$ & VLA:C \\ [0.1ex]
$221$ & $30$ & Ka & $8.6 \pm 0.9$ & $0.06$ & VLA:C \\ [0.1ex]
$221$ & $32$ & Ka & $7.7 \pm 0.8$ & $0.06$ & VLA:C \\ [0.1ex]
$221$ & $34$ & Ka & $6.8 \pm 0.7$ & $0.06$ & VLA:C \\ [0.1ex]
$221$ & $36$ & Ka & $6.0 \pm 0.6$ & $0.06$ & VLA:C \\ [0.1ex]
$392$ & $2.2$ & S & $<0.12$ & $0.04$ & VLA:B \\ [0.1ex]
$392$ & $2.8$ & S & $0.34 \pm 0.05$ & $0.03$ & VLA:B \\ [0.1ex]
$392$ & $3.2$ & S & $0.55 \pm 0.06$ & $0.02$ & VLA:B \\ [0.1ex]
$392$ & $3.8$ & S & $0.88 \pm 0.09$ & $0.02$ & VLA:B \\ [0.1ex]
$392$ & $4.5$ & C & $1.5 \pm 0.2$ & $0.02$ & VLA:B \\ [0.1ex]
$392$ & $5.5$ & C & $2.6 \pm 0.3$ & $0.02$ & VLA:B \\ [0.1ex]
$392$ & $6.5$ & C & $3.9 \pm 0.4$ & $0.02$ & VLA:B \\ [0.1ex]
$392$ & $7.5$ & C & $5.0 \pm 0.5$ & $0.02$ & VLA:B \\ [0.1ex]
$392$ & $8.5$ & X & $6.1 \pm 0.6$ & $0.02$ & VLA:B \\ [0.1ex]
$392$ & $9.5$ & X & $6.8 \pm 0.7$ & $0.02$ & VLA:B \\ [0.1ex]
$392$ & $10.5$ & X & $7.3 \pm 0.7$ & $0.02$ & VLA:B \\ [0.1ex]
$392$ & $11.5$ & X & $7.6 \pm 0.8$ & $0.02$ & VLA:B \\ [0.1ex]
$392$ & $13$ & Ku & $7.8 \pm 0.8$ & $0.02$ & VLA:B \\ [0.1ex]
$392$ & $15$ & Ku & $7.8 \pm 0.8$ & $0.02$ & VLA:B \\ [0.1ex]
$392$ & $17$ & Ku & $7.6 \pm 0.8$ & $0.02$ & VLA:B \\ [0.1ex]
$392$ & $19$ & K & $7.5 \pm 0.8$ & $0.03$ & VLA:B \\ [0.1ex]
$392$ & $21$ & K & $7.2 \pm 0.7$ & $0.03$ & VLA:B \\ [0.1ex]
$392$ & $23$ & K & $6.9 \pm 0.7$ & $0.04$ & VLA:B \\ [0.1ex]
$392$ & $25$ & K & $6.4 \pm 0.6$ & $0.03$ & VLA:B \\ [0.1ex]
$392$ & $30$ & Ka & $6.0 \pm 0.6$ & $0.02$ & VLA:B \\ [0.1ex]
$392$ & $32$ & Ka & $5.7 \pm 0.6$ & $0.03$ & VLA:B \\ [0.1ex]
$392$ & $34$ & Ka & $5.5 \pm 0.6$ & $0.03$ & VLA:B \\ [0.1ex]
$392$ & $36$ & Ka & $5.2 \pm 0.5$ & $0.02$ & VLA:B \\ [0.1ex]
$558$ & $2.2$ & S & $1.4 \pm 0.3$ & $0.2$ & VLA:A $\rightarrow$ D \\ [0.1ex]
$557$ & $2.8$ & S & $1.6 \pm 0.2$ & $0.1$ & VLA:A $\rightarrow$ D \\ [0.1ex]
$557$ & $3.2$ & S & $2.2 \pm 0.3$ & $0.1$ & VLA:A $\rightarrow$ D \\ [0.1ex]
$557$ & $3.8$ & S & $2.6 \pm 0.3$ & $0.06$ & VLA:A $\rightarrow$ D \\ [0.1ex]
$557$ & $4.5$ & C & $3.8 \pm 0.4$ & $0.02$ & VLA:A $\rightarrow$ D \\ [0.1ex]
$557$ & $5.5$ & C & $5.0 \pm 0.5$ & $0.02$ & VLA:A $\rightarrow$ D \\ [0.1ex]
$557$ & $6.5$ & C & $6.1 \pm 0.6$ & $0.03$ & VLA:A $\rightarrow$ D \\ [0.1ex]
$557$ & $7.5$ & C & $6.5 \pm 0.7$ & $0.03$ & VLA:A $\rightarrow$ D \\ [0.1ex]
$557$ & $8.5$ & X & $7.0 \pm 0.7$ & $0.02$ & VLA:A $\rightarrow$ D \\ [0.1ex]
$557$ & $9.5$ & X & $7.2 \pm 0.7$ & $0.02$ & VLA:A $\rightarrow$ D \\ [0.1ex]
$557$ & $10.5$ & X & $7.2 \pm 0.7$ & $0.02$ & VLA:A $\rightarrow$ D \\ [0.1ex]
$557$ & $11.5$ & X & $7.0 \pm 0.7$ & $0.03$ & VLA:A $\rightarrow$ D \\ [0.1ex]
$557$ & $13$ & Ku & $6.9 \pm 0.7$ & $0.02$ & VLA:A $\rightarrow$ D \\ [0.1ex]
$557$ & $15$ & Ku & $6.7 \pm 0.7$ & $0.02$ & VLA:A $\rightarrow$ D \\ [0.1ex]
$557$ & $17$ & Ku & $6.3 \pm 0.6$ & $0.02$ & VLA:A $\rightarrow$ D \\ [0.1ex]
$557$ & $19$ & K & $5.7 \pm 0.6$ & $0.03$ & VLA:A $\rightarrow$ D \\ [0.1ex]
$557$ & $21$ & K & $5.2 \pm 0.5$ & $0.03$ & VLA:A $\rightarrow$ D \\ [0.1ex]
$557$ & $23$ & K & $4.8 \pm 0.5$ & $0.02$ & VLA:A $\rightarrow$ D \\ [0.1ex]
$557$ & $25$ & K & $4.5 \pm 0.5$ & $0.03$ & VLA:A $\rightarrow$ D \\ [0.1ex]
$557$ & $30$ & Ka & $3.6 \pm 0.4$ & $0.05$ & VLA:A $\rightarrow$ D \\ [0.1ex]
$557$ & $32$ & Ka & $3.5 \pm 0.4$ & $0.04$ & VLA:A $\rightarrow$ D \\ [0.1ex]
$557$ & $34$ & Ka & $3.7 \pm 0.4$ & $0.09$ & VLA:A $\rightarrow$ D \\ [0.1ex]
$557$ & $36$ & Ka & $3.5 \pm 0.4$ & $0.06$ & VLA:A $\rightarrow$ D \\ [0.1ex]
$572$ & $15.5$ & Ku & $7.0 \pm 0.7$ & $0.06$ & AMI-LA \\ [0.1ex]
$596$ & $15.5$ & Ku & $7.2 \pm 0.7$ & $0.04$ & AMI-LA \\ [0.1ex]
$635$ & $15.5$ & Ku & $6.9 \pm 0.7$ & $0.05$ & AMI-LA \\ [0.1ex]
$672$ & $15.5$ & Ku & $5.9 \pm 0.6$ & $0.1$ & AMI-LA \\ [0.1ex]
$674$ & $15.5$ & Ku & $5.9 \pm 0.6$ & $0.05$ & AMI-LA \\ [0.1ex]
$704$ & $15.5$ & Ku & $6.2 \pm 0.6$ & $0.05$ & AMI-LA \\ [0.1ex]
$777$ & $15.5$ & Ku & $5.6 \pm 0.6$ & $0.05$ & AMI-LA \\ [0.1ex]
$850$ & $15.5$ & Ku & $4.5 \pm 0.5$ & $0.1$ & AMI-LA \\ [0.1ex]
$881$ & $15.5$ & Ku & $5.3 \pm 0.5$ & $0.04$ & AMI-LA \\ [0.1ex]
$940$ & $15.5$ & Ku & $4.7 \pm 0.5$ & $0.03$ & AMI-LA \\ [0.1ex]
$1009$ & $15.5$ & Ku & $4.3 \pm 0.4$ & $0.03$ & AMI-LA \\ [0.1ex]
$1121$ & $15.5$ & Ku & $4.4 \pm 0.4$ & $0.03$ & AMI-LA \\ [0.1ex]
$1136$ & $15.5$ & Ku & $4.7 \pm 0.5$ & $0.03$ & AMI-LA \\ [0.1ex]
$1143$ & $15.5$ & Ku & $4.6 \pm 0.5$ & $0.03$ & AMI-LA \\ [0.1ex]
$1148$ & $15.5$ & Ku & $4.8 \pm 0.5$ & $0.03$ & AMI-LA \\ [0.1ex]
$1149$ & $15.5$ & Ku & $4.5 \pm 0.5$ & $0.08$ & AMI-LA \\ [0.1ex]
$1171$ & $15.5$ & Ku & $4.4 \pm 0.4$ & $0.03$ & AMI-LA \\ [0.1ex]
$1192$ & $15.5$ & Ku & $4.4 \pm 0.4$ & $0.05$ & AMI-LA \\ [0.1ex]
$1232$ & $15.5$ & Ku & $3.6 \pm 0.4$ & $0.1$ & AMI-LA \\ [0.1ex]
$1300$ & $15.5$ & Ku & $4.1 \pm 0.4$ & $0.08$ & AMI-LA \\ [0.1ex]
$1358$ & $15.5$ & Ku & $3.5 \pm 0.4$ & $0.03$ & AMI-LA \\ [0.1ex]
$1425$ & $15.5$ & Ku & $3.3 \pm 0.3$ & $0.04$ & AMI-LA \\ [0.1ex]
\hline
\end{longtable}
\tablefoot{\footnotesize Radio observations of SN\,2019oys conducted with the AMI-LA and the VLA. $\Delta t$ is the time since optical discovery. $\nu$ is the observed frequency in GHz.}

\begin{longtable}{ccccccccc}
\caption{SN\,2019oys - fitting spectrum on $\Delta t = 201$ and $221$ days \label{tab:fitting}}\\
\hline\hline
Model & $\rm \Delta t$ & $F_{\rm \nu,a}$ & $\nu_{\rm a}$ & $\rm \beta$ & $\rm \nu_{\rm ff}$ & $\chi^2$ (DOF) & p-value & BIC \\
- & [Day] & [mJy] & [GHz] & - & [GHz] & - & - & - \\
\hline
\endhead
\hline
\endfoot
SSA & $201$ & $17.8 \pm 0.5$ & $22.8 \pm 0.6$ & $2.0^{a}$ & - & 25.5 (15) & 0.06 & 31.3 \\ [0.1ex]
SSA & $221$ & $13.5 \pm 0.6$ & $19.7 \pm 0.5$ & $2.0 \pm 0.2$ & - & 32.8 (15) & 0.005 & 41.6 \\ [0.1ex]
SSA + FFA & $201$ & $17.4 \pm 0.5$ & $21.5 \pm 0.4$ & $1.7^{a}$ & $3.1 \pm 0.4$ & 1.68 (14) & 0.8 & 19.2 \\ [0.1ex]
SSA + FFA & $221$ & $13.6 \pm 0.7$ & $17.9 \pm 0.6$ & $1.7 \pm 0.2$ & $3.5 \pm 0.4$ & 8.12 (14) & 0.89 & 19.8 \\ [0.1ex]
\end{longtable}
\tablefoot{\footnotesize Parameters of the fitting processes of the spectrum of SN\,2019oys on $\Delta t = 201$ and $221$ days to Eq. \ref{eq: parameterized} with and without additional FFA. The fitted parameters are the radio spectral peak, $F_{\rm \nu,a}$, its frequency, $\nu_{\rm a}$, the optically thin spectral slope, $\beta$. $\Delta t = 201$ days are marked with $^{a}$ as we used the $\beta$ fitted for $\Delta t = 221$ days (due to the lack of spectral coverage). When fitting FFA as an additional absorption mechanism we also fit for the FFA frequency, $\nu_{\rm ff}$. DOF is the number of degrees of freedom.}

\newpage
\section{Spectra corner plots}
\label{sec: corner_plots}

\begin{figure}[!ht]
\includegraphics[width=1\linewidth]{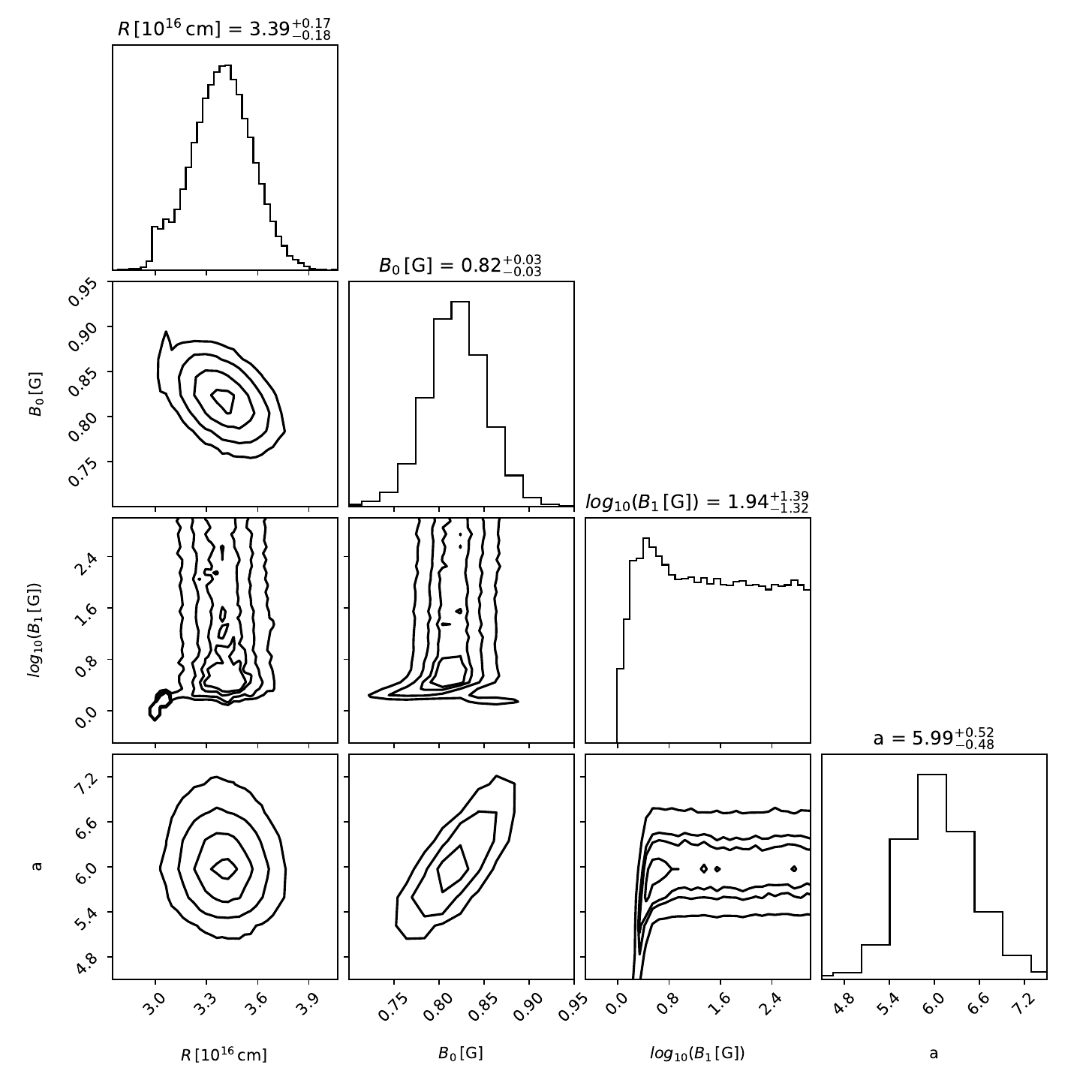}
\caption{\footnotesize{Posterior probability distributions from our MCMC analysis of the radio spectrum obtained $392$ days after discovery according to the CSM inhomogeneities model introduced in \S\ref{sec: inhomogen} and in \cite{Bjornsson_2013}. The data clearly does not manage to constrain the upper value of the magnetic field distribution, $B_1$, which is uniquely defined by the transition from the spectral peak to the optically thin regime.  \label{fig: corner_392_bjor}}}
\end{figure}

\begin{figure}
\includegraphics[width=1\linewidth]{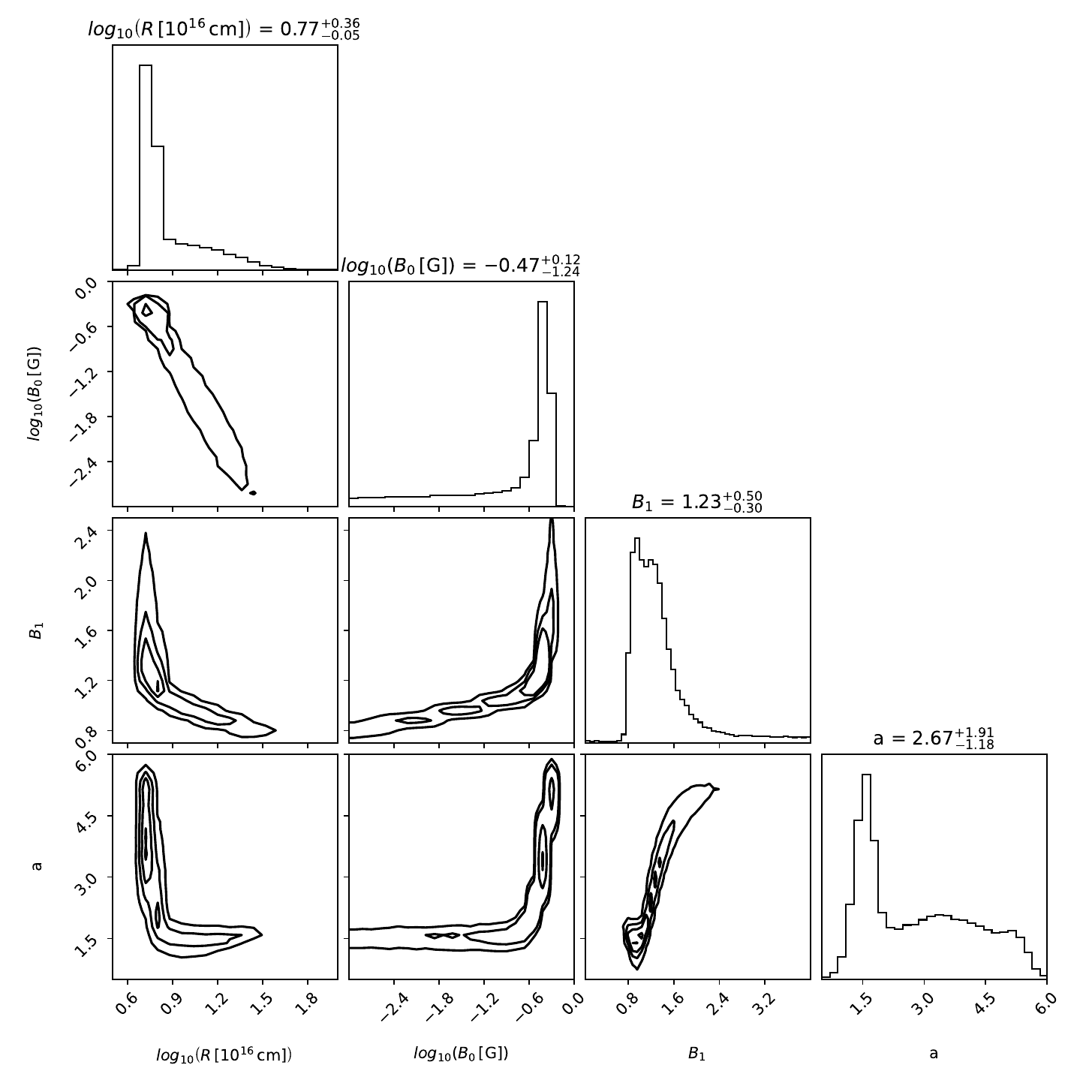}
\caption{\footnotesize{Posterior probability distributions from our MCMC analysis of the radio spectrum obtained $557$ days after discovery according to the CSM inhomogeneities model introduced in \S\ref{sec: inhomogen} and in \cite{Bjornsson_2013}. The data does not manage to constrain the lower value of the magnetic field distribution, $B_0$, which is uniquely defined by the transition from the optically thick regime to the spectral peak.  \label{fig: corner_557_bjor}}}
\end{figure}

\end{appendix}
\end{document}